\documentclass[11pt,preprint]{aastex}
\usepackage{natbib}
\usepackage{iondefs}

\begin{document}

\title{Limits on Reddening and Gas-to-Dust Ratios for Seven \\ Intermediate Redshift Damped {\Lya} Absorbers \\ from Diffuse Interstellar Bands}

\author{\sc 
Brandon Lawton\altaffilmark{1}, 
Christopher W. Churchill\altaffilmark{1}, 
Brian A. York\altaffilmark{2}, 
Sara L. Ellison\altaffilmark{2}, \\ 
Theodore P. Snow\altaffilmark{3}, 
Rachel A. Johnson\altaffilmark{4},
Sean G. Ryan\altaffilmark{5}, and
Chris R. Benn\altaffilmark{6}
}

\altaffiltext{1}{Department of Astronomy, MSC 4500, New Mexico State
University, P.O. Box 30001, Las Cruces, NM 88003; blawton@nmsu.edu,
cwc@nmsu.edu}

\altaffiltext{2}{Department of Physics \& Astronomy, University of
Victoria, 3800 Finnerty Rd., Victoria, V8W 1A1, British Columbia,
Canada; briany@uvic.ca, sarae@uvic.ca}

\altaffiltext{2}{Center for Astrophysics and Space Astronomy,
University of Colorado at Boulder, 389 UCB, Boulder, CO 80309;
Theodore.Snow@colorado.edu}

\altaffiltext{4}{Oxford Astrophysics, Denys Wilkinson Building, Keble
Road, Oxford OX1 3RH, UK; raj@astro.ox.ac.uk}

\altaffiltext{5}{Centre for Astrophysics Research, University of
Hertfordshire, College Lane, Hatfield AL10 9AB, UK;
s.g.ryan@herts.ac.uk}

\altaffiltext{6}{Isaac Newton Group, Apartado 321, E-38700 Santa Cruz
de La Palma, Spain; crb@ing.iac.es}

\begin{abstract}

We present equivalent width measurements and limits of six diffuse
interstellar bands (DIBs, $\lambda$4428, $\lambda$5705, $\lambda$5780,
$\lambda$5797, $\lambda$6284, and $\lambda$6613) in seven damped
{\Lya} absorbers (DLAs) over the redshift range $0.091 \leq z \leq
0.524$, sampling $20.3 \leq \log N({\HI}) \leq 21.7$. DIBs were
detected in only one of the seven DLAs, that which has the highest
reddening and metallicity.  Based upon the Galactic DIB--$N({\HI})$
relation, the $\lambda$6284 DIB equivalent width upper limits in four
of the seven DLAs are a factor of 4-10 times below the $\lambda$6284
DIB equivalent widths observed in the Milky Way, but are not
inconsistent with those present in the Magellanic Clouds. Assuming the
Galactic DIB--$E(B-V)$ relation, we determine reddening upper limits
for the DLAs in our sample.  Based upon the $E(B-V)$ limits, the
gas-to-dust ratios, $N({\HI})/E(B-V)$, of the four aforementioned DLAs
are at least $\sim 5$ times higher than that of the Milky Way ISM.
The ratios of two other DLAs are at least a factor of a few times
higher.  The best constraints on reddening derive from the upper
limits for the $\lambda$5780 and $\lambda$6284 DIBs, which yield
$E(B-V) \leq 0.08$ for four of the seven DLAs.  Our results suggest
that, in DLAs, quantities related to dust, such as reddening and
metallicity, appear to have a greater impact on DIB strengths than
does {\HI} gas abundance; the organic molecules likely responsible for
DIBs in DLA selected sightlines are underabundant relative to
sightlines in the Galaxy of similarly high $N({\HI})$.  With regards
to the study of astrobiology, this could have implications for the
abundance of organic molecules in redshifted galaxies.  However, since
DLAs are observed to have low reddening, selection bias likely plays a
role in the apparent underabundance of DIBs in DLAs.

\end{abstract}

\keywords{
dust, extinction --
galaxies: ISM and abundances --
quasars: absorption lines and individual (AO 0235+164, Q0738+313, B2
0827+243, PKS 0952+179, PKS 1127--145, Q1229--020) --
techniques: spectroscopic
}

\section{Introduction}

Since their discovery by \citet{hege22}, several hundred diffuse
interstellar bands (DIBs) have been studied
\citep{gala00,jenn94,tuai00,wese00,hobbs07}, and yet no positive
identifications of the carriers have been made.  The DIBs span the
visible spectrum between 4000 and 13,000~{\AA}.  Despite no positive
identifications, several likely organic molecular candidates have
emerged as the sources of the DIBs, including polycyclic aromatic
hydrocarbons (PAHs), fullerenes, long carbon chains, and polycyclic
aromatic nitrogen heterocycles (PANHs) \citep[e.g.,][]{herb95, snow01,
cox06a, hudg05}.  The organic-molecular origin of the DIBs may give
them an importance to astrobiology; PAHs are now considered an
important early constituent to the inventory of organic compounds on
Earth \citep{bada02}.  Via their infrared emitting vibrational bands,
PAHs have been observed in high redshift dusty ultra-luminous infrared
galaxies \citep[ULIRGs, e.g.,][]{yan05}.  Searching for DIBs using the
technique of quasar absorption lines provides a different approach for
charting the presence of possible organics to high redshift.  As such,
observing DIBs in high redshift galaxies may offer an independent
method for constraining the environmental conditions in early-epoch
galaxies governing the abundances of organic molecules, determining
the cosmic epoch at which organic molecules first formed, and
ultimately charting their evolution with redshift.

Aside from the hundreds of detections within the Galaxy
\citep[e.g.][]{gala00,jenn94,tuai00,wese00,hobbs07}, DIBs have been
detected in the Magellanic Clouds \citep{welt06,cox06b,cox07}, M31
\citep{cord08}, seven starburst galaxies \citep{heck00}, active galaxy
Centaurus A via supernova 1986A \citep{rich87}, spiral galaxy NGC 1448
via supernovae 2001el and 2003hn \citep{soll05}, one damped {\Lya}
absorber (DLA) at $z=0.524$ toward the quasar (QSO) AO~0235+164
\citep{junk04, lawt06}, and one $z=0.157$ {\CaII} selected absorber
toward QSO J0013--0024 \citep{elli07}.

There are several environmental factors, such as {\HI} column density
\citep{herb95,welt06}, reddening \citep{welt06}, and metallicity and
ionizing radiation \citep{cox07}, that are related to DIB strengths.
In the Galaxy, DIB absorption strengths correlate strongly with
$N({\HI})$ \citep{herb95,welt06}.  However, in the Magellanic Clouds,
DIBs are weaker by factors of 7-9 (LMC) and $\sim20$ (SMC) compared to
those observed in the Galaxy with similar $N({\HI})$ \citep{welt06}.
It is not known whether other galaxies in the Local Group and beyond
obey the Galactic DIB--$N({\HI})$ relation, or, like the Magellanic
Clouds, show departures from this relation.  Galaxies with high
$N({\HI})$ observed in absorption (i.e., DLAs) that reside at low to
intermediate redshifts (where the prominent DIBs fall in the optical
region) provide excellent astrophysical laboratories with which to
investigate this issue.

In this paper we search for $\lambda$4428, $\lambda$5780,
$\lambda$5797, $\lambda$6284, and $\lambda$6613 DIB absorption in
seven low to intermediate redshift DLAs.  In \S~\ref{sec:sample}, we
give a brief summary of each intervening DLA in our sample.  In
\S~\ref{sec:obs}, we discuss the spectroscopic observations and data
reduction of the background QSOs.  In \S~\ref{sec:analysis}, we
explain the procedure of our analysis, and the resulting spectra.  In
\S~\ref{sec:results}, we present our results and compare our data to
the Galactic DIB--$N({\HI})$ relation, deduce upper limits for the
reddening, $E(B-V)$, determine lower limits on the gas-to-dust ratios,
and discuss the role of metallicity for our sample of DLAs.  We
conclude in \S~\ref{sec:conclusions}.

\section{DLA Sample}
\label{sec:sample}

To potentially maximize our chances of detecting DIBs, and to test the
Galactic DIB--$N({\HI})$ relation in redshifted galaxies, we selected
DLAs toward background QSOs having the highest $N({\HI})$ in
absorption.  We limited the redshift range such that the strongest
DIBs would fall in the optical region.  Thus, we selected the highest
$N({\HI})$ DLA galaxies in the redshift range $0.09<z<0.52$.  DLAs, by
definition, have a large neutral gas column density ($N({\HI})\geq 2
\times 10^{20}$ cm$^{-2}$).  However, they are observed to have low
reddening ($E(B-V)<0.04$) \citep{elli05} and low metallicity
(typically $Z\sim0.1 Z_{\odot}$).  The low metallicity and reddening
of DLAs suggest that the gaseous environments selected by DLA
absorption differ from the Galactic ISM in which DIBs are observed.
As we discuss in \S~\ref{sec:reddening}, the low reddening of DLAs is
not merely a bias arising from the necessity of low extinction if the
background quasar is to be detected at all, since DLAs in
radio-selected quasars have similarly low reddenings.

Below, we describe the basic properties of each of the seven DLAs
comprising our sample.  Table~\ref{tab:DLAs} summarizes the main
parameters, including abundance information.  The columns, from left
to right, list the QSO, absorption redshift of the DLA, column density
of neutral hydrogen, the zinc abundance, the iron abundance, and the
associated references.  The {\Lya} line is used to measure the DLA
redshift.  All impact parameters have been converted to a $\Lambda$CDM
cosmology ($H_0=70$~km~s$^{-1}$~Mpc$^{-1}$, $\Omega_{m}=0.3$,
$\Omega_{\Lambda}=0.7$). 

The $z_{abs}=0.524$ DLA ({\#1}) toward AO~0235+164 is unique in our
sample for many reasons. It is the only DLA known to have DIB
absorption\footnote{The DIB-bearing {\CaII} absorber reported by
\citet{elli07} is likely to be a DLA.  However, this cannot be
confirmed without a spectrum of the {\Lya} absorption.}, the $\lambda
4428$ DIB \citep{junk04} and the $\lambda 5705$ and $\lambda 5780$
DIBs \citep{lawt06}.  The $\lambda 5797$, $\lambda 6284$, and $\lambda
6613$ DIBs have limits that are not as sensitive due to sky lines.
The associated optical galaxy with the smallest impact parameter,
$6.7~h_{70}^{-1}$~kpc, is a late-type spiral and is assumed to be the
absorber \citep{raot03,burb96,yann89}.  However, \citet{chen03} argue
that many of the galaxies in the optical field have the same redshift
and may collectively be responsible for the DLA absorption.  The
adopted neutral hydrogen column density is from \citet{junk04} and is
consistent with $N(\HI)=(4.5\pm0.4)\times 10^{21}$~{\cmsq} from
\citet{turn03}.  The measured metallicities \citep{turn03,junk04}
assume solar abundance ratios and solar metallicity for the foreground
Galactic absorption.  \citet{junk04} state that the difference between
their metallicities and those of \citet{turn03} are likely due to
different realizations of the noise in the X-ray data and/or the
variability \citep{riek76} of this QSO.  Despite the differences in
these measurements, this DLA has the highest metallicity in our
sample.  The DLA also has significant reddening with $E(B-V)=0.23$,
$R_{v}=2.5$, and a strong graphitic dust feature at 2175~{\AA}.

The $z=0.091$ DLA ({\#2}) toward Q0738+313 (OI363) is probably a low
surface brightness galaxy with an impact parameter of
$<3.5~h_{70}^{-1}$~kpc \citep{turn01}.  This DLA is one of two found
along the QSO sightline of Q0738+313 and was first reported by
\citet{raot98}.

The other DLA, $z_{abs}=0.221$ ({\#3}), toward the QSO Q0738+313 is
probably a dwarf spiral with an impact parameter of
$20.5~h_{70}^{-1}$~kpc \citep{turn01}.

The $z_{abs}=0.518$ DLA ({\#4}) toward B2 0827+243 is likely a
disturbed spiral galaxy with extended gas that produces the observed
hydrogen absorption at an impact parameter of $38.2~h_{70}^{-1}$~kpc
\citep{raot03}.  \citet{khare04} note that it appears the DLA requires
a significant radiation field, similar to the radiation within the
dense clouds in our Galaxy, to create a low $N(\FeI)/N(\FeII)$ upper
limit of 10$^{-3}$.

The absorbing galaxy giving rise to the $z_{abs}=0.239$ DLA ({\#5})
toward PKS~0952+179 has not been confirmed.  However, two candidates
lie within $<4.6~h_{70}^{-1}$~kpc \citep{raot03}.  The authors note
that these two galaxies appear to be nearly edge-on and are classified
as dwarf low surface brightness galaxies. The DLA appears to have
multiple {\Lya} components around the central line at $z=0.239$
\citep{raot00}.  The adopted hydrogen column density is measured from
the central {\Lya} feature.  However, the velocity structure suggests
a possible clustering.

The $z_{abs}=0.313$ DLA ({\#6}) toward PKS~1127--145 is possibly the
remnant of a dwarf low surface brightness galaxy tidally disturbed by
more massive spiral galaxies in the same field \citep{raot03}.  The
absorber appears to overlap the QSO point spread function, thus,
\citet{raot03} find an upper limit for the impact parameter of
$6.9~h_{70}^{-1}$~kpc.  \citet{turn03} discuss the difficulty of
determining a metallicity for this system due to uncertainties in
their X-ray spectrum.  We have estimated a lower limit of the iron
abundance from a VLT/UVES spectrum kindly donated for this work by
Dr.\ M. Murphy.  The system has a complicated velocity structure.
However, fixing the Doppler parameter, redshift, and number of Voigt
profiles yields a lower limit on the column density of $\log
N(\FeII)>15.7$~atoms~cm$^{-2}$.  Repeating this analysis using the
apparent optical depth \citep{sava91} gives a slightly more
conservative limit of $\log N(\FeII)>15.2$~atoms~cm$^{-2}$, which we
adopt for the column density of iron.  Assuming the \citet{lodd03}
solar abundances, we deduce a lower limit of $\hbox{[Fe/H]}>-2$.

The absorbing galaxy responsible for the $z_{abs}=0.395$ DLA ({\#7})
along the Q1229--029 sight--line is a low surface brightness galaxy
with an impact parameter of $8.2~h_{70}^{-1}$~kpc \citep{lebr97,
stei94}.

\section{Observations and Data Reduction}
\label{sec:obs}

Observations of the seven DLAs were obtained with seven facilities
toward six QSO sightlines between July 2002 and October 2005.  The S/N
of the QSO spectra range from 5--150~pixel$^{-1}$.  All quoted S/N
measurements are calculated in the regions of the expected locations
of the redshifted $\lambda4428$, $\lambda5780$, $\lambda5797$,
$\lambda6284$, and $\lambda6613$ DIBs.  The journal of observations is
presented in Table~\ref{tab:obs}.  Cataloged are the QSO, facility and
instrument used in the observation, the grating/grism, the slit width,
the UT date of the observation, the total exposure time in seconds,
and the wavelength coverage of each spectrum in angstroms.

\subsection{Observations}

QSO spectra covering the $z=0.524$, $z=0.239$, and $z=0.395$ DLAs
along the sightlines AO~0235+164, PKS~0952+179, and Q1229--020 were
obtained with the FORS2 spectrograph on the Very Large Telescope
(VLT).  All resolutions are obtained by measuring unresolved sky
emission lines.  Observations of A0~0235+164 and Q1229--020 use the
same grating and the resolutions agree within reasonable
uncertainties.

A QSO spectrum covering the $z=0.091$ and $z=0.221$ DLAs along the
Q0738+313 sightline was acquired with the DIS spectrograph on the
Apache Point Observatory (APO) 3.5~m telescope.  DIS is configured
with a dichroic that splits the light to a blue chip and a red chip at
${\sim}5500$~{\AA}.  The resolution is measured directly from
unresolved sky emission lines.

A Keck/HIRES spectrum of B2 0827+243 covering the DLA at $z=0.518$ was
kindly provided by Dr.\ W. Sargent.  The resolution is measured by
unresolved atmospheric absorption lines (we did not have access to sky
emission line data for this object).

QSO spectra covering the $z=0.239$ and $z=0.395$ DLAs along the
sightlines toward PKS~0952+179 and Q1229--020 were obtained using the
ISIS spectrograph on the 4~m William Herschel Telescope (WHT).  The
resolutions were measured directly from unresolved sky emission lines;
the uncertainties are relatively large because of the low S/N in these
data.

A UVES/VLT spectrum covering the $z=0.313$ DLA toward PKS~1127--145
was kindly provided by Dr.\ M. Murphy.  The resolution is taken from
\citet{dekk00}.  We did not have access to sky data for this spectrum
nor were there any unresolved atmospheric absorption lines, so we
could not estimate the resolution directly.

A QSO spectrum covering the $z=0.313$ DLA along the sightline toward
PKS~1127--145 was obtained with the GMOS spectrograph on the 8.1~m
Gemini South telescope.  The resolution is taken from the online
Gemini Science Operations GMOS Instrument Manual.  We had limited sky
data, so we did not estimate the resolution directly.  However, due to
large equivalent width limits of the GMOS data, the effect of
uncertainties in resolution should be minimal and not affect results
in this paper (see Table~\ref{tab:EWs}).

\subsection{Data Reduction}

With the exception of the VLT/UVES and Gemini/GMOS spectra of
PKS~1127--145, the data were reduced using standard IRAF\footnote{IRAF
is distributed by the National Optical Astronomy Observatories, which
are operated by the Association of Universities for Research in
Astronomy, Inc., under cooperative agreement with the National Science
Foundation.} packages.  The IRAF reduction process involved bias
subtraction, flat-fielding, spectrum extraction, and wavelength
calibration using standard lamps.  Once wavelength calibrated, each
spectrum was continuum fit manually using \textit{sfit} to achieve the
lowest residuals in the regions of the DIBs where no detections are
observed.  Near telluric lines or problematic sky subtraction, the
continuum was fit using regions bracketing these features.  The flux,
uncertainty, sky (when acquired), and continuum spectra are normalized
and then the individual spectra are optimally combined (using an
algorithm of our own design).  For the Gemini/GMOS-S spectrum, data
reduction was performed using the IRAF Gemini tools in the
\textit{gmos} package.  The IRAF task, \textit{telluric}, was also
used on the Gemini/GMOS-S spectrum \citep{lawt06}.  The UVES spectrum
was reduced using the standard ESO pipeline and a custom code called
the UVES Post--Pipeline Echelle Reduction \citep[{\sc uves
popler},][]{popler}.

\section{Data Analysis and Spectra}

\label{sec:analysis}

\subsection{Analysis of DIBs}

A modified method of \citet{schn93} was developed and used to search
for the presence of $\lambda$4428, $\lambda$5780, $\lambda$5797,
$\lambda$6284, and $\lambda$6613 DIB absorption.  The Schneider et
al.\ technique is optimized for objectively locating unresolved
features in spectra.  However, some DIBs are resolved in our spectra.
Thus, we modified the \citet{schn93} method to be optimized for both
unresolved and resolved features by combining the DIB's intrinsic full
width at half maximum (\hbox{\small FWHM}) with the instrumental
spread function (ISF).  The procedure is outlined in Appendix A.1.
The calculation used by \citet{schn93} (given in
Eq.~\ref{EQ:dpsewlim}) transforms to Eq.~\ref{EQ:EWlimit}.

\begin{figure}
\includegraphics[angle=-90,width=1.0\textwidth]{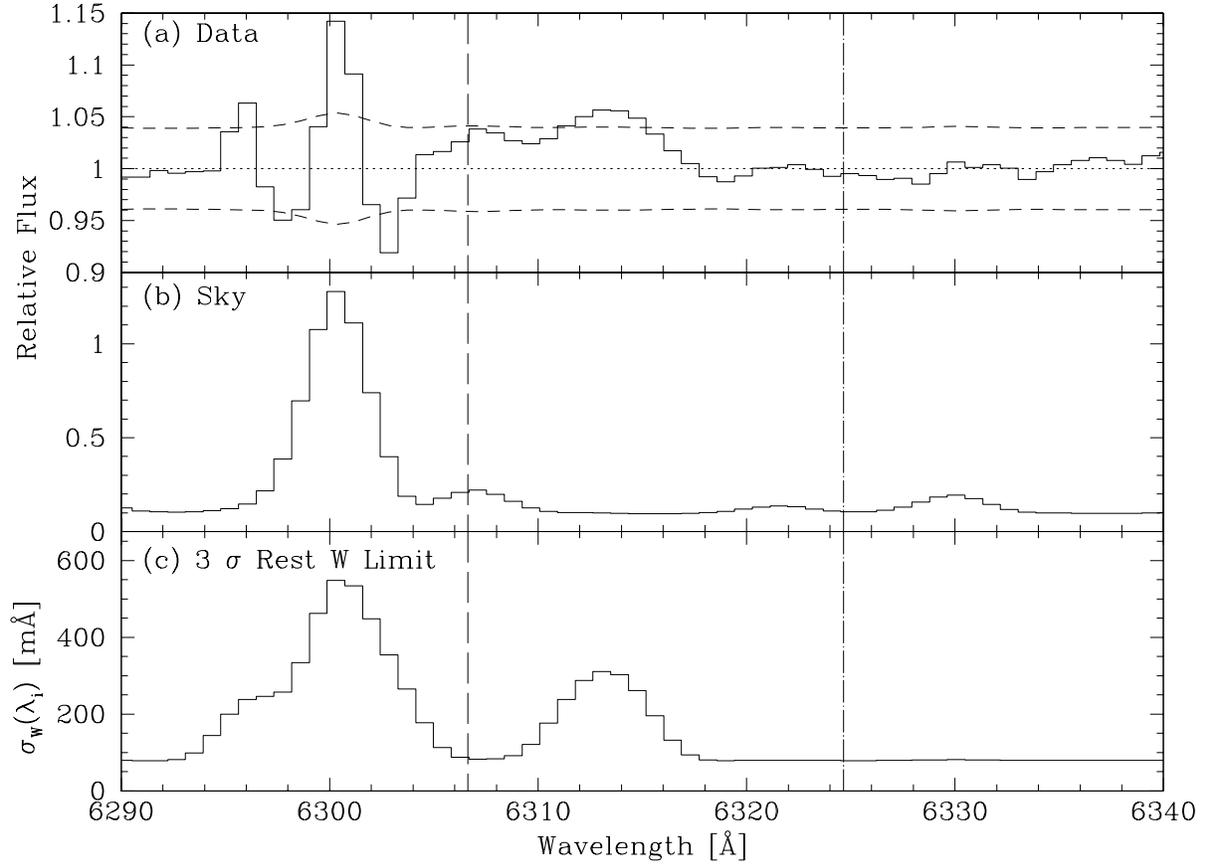}
\caption{Equivalent width limit analysis of the Q0738+313 $z=0.091$
DLA. ---($a$) Relative flux in the region of the $\lambda$5780 and
$\lambda$5797 DIBs (vertical long dashed line and vertical long
dashed-dotted line, respectively).  The histogram is the relative flux
data, and the short dashed lines are the associated $\pm3~\sigma$ flux
errors.  ---($b$) Relative sky flux.  ---($c$) The $3~\sigma$ rest
equivalent width limits in m{\AA}.  The two peaks in panel ($c$) are
due to the problematic sky subtraction from the [OI] sky emission line
at $\lambda\sim 6300$~{\AA} and the poor continuum fit at $\lambda\sim
6314$~{\AA} where the residuals are used, as explained in
Eq.~\ref{EQ:flux_error}.\label{p:kspace}}
\end{figure}

In addition, we replaced the normalized flux error of \citet{schn93}
by the residuals of the data in pixels where the normalized flux
deviates significantly from the continuum (see
Eqs.~\ref{EQ:flux_error}, \ref{EQ:residual}, and
\ref{EQ:norm_flux_error}).  This results in a more conservative
detection threshold (equivalent width limit) in the cases of
problematic sky subtraction or continuum fits.  As an example of this
method, Fig.~\ref{p:kspace}$a$ displays the relative flux of Q0738+313
in the region of the expected positions of the $\lambda$5780 and
$\lambda$5797 DIBs for the $z=0.091$ DLA.  Fig.~\ref{p:kspace}$b$
contains the sky data.  Fig.~\ref{p:kspace}$c$ contains the associated
3~$\sigma$ rest-frame detection thresholds for a Gaussian profile with
the expected \hbox{\small FWHM} of the redshifted $\lambda$5780 DIB.
Both the $\lambda$5780 and $\lambda$5797 DIBs are unresolved.
However, if they had been resolved they would have yielded different
equivalent width limits because the limits depends upon the redshifted
\hbox{\small FWHM} for resolved DIBs.  The problematic sky subtraction
at $6300$~{\AA} and the problematic continuum fit at $6314$~{\AA}
result in conservative limits due to large residuals (see
Eqs.~\ref{EQ:flux_error}, \ref{EQ:norm_flux_error}, and
\ref{EQ:EWlimit}).  The region around the $\lambda$5797 DIB is
``clean'' in that no residuals are used in the equivalent width limit
calculations.  The equivalent width limits of the DIBs in the
AO~0235+164 DLA measured by \citet{lawt06} are less stringent but not
inconsistent with those measured by this method.

Possible detections were visually inspected to determine if they were
a DIB or perhaps another feature or sky line residual.  Checking each
candidate feature is essential because of a small uncertainty in the
rest wavelength of each DIB as well as a small uncertainty in the
redshift of DIB absorbing gas relative to the {\Lya} determined
redshift of the DLA.  For most DIBs, we were able to measure only
equivalent width limits.

To quantify confidence levels in the measured detection thresholds
determined with our modified method, we have estimated the $1~\sigma$
uncertainties in the measured equivalent width limits.  There is some
uncertainty in the resolution (ISF) of each spectrum.  We estimate the
uncertainty in the resolution as the standard deviation of the
$\lambda/\hbox{\small FWHM}$ ratios of unresolved sky absorption lines
or emission lines, where the $\lambda$ are the line centers.  If sky
data are not available, we excluded the uncertainty in resolution.  In
addition, there is uncertainty in the accuracy of the continuum fits,
which we estimated using the technique of \citet{semb91}.  Of
importance, is the fact that DIBs have measured uncertainties in their
rest-frame wavelengths and \hbox{\small FWHMs} \citep{jenn94}.  The
uncertainty in the equivalent width limit due to uncertainty in the
wavelength is the standard deviation of the individual equivalent
width limits computed over the range of the redshifted DIB wavelength
uncertainty.  The full explanation of ``error'' propagation to obtain
uncertainties in the equivalent width limits is given in Appendix A.2
(see Eq.~\ref{EQ:var}).

Contained within Fig.~\ref{p:kspace} is an example of our equivalent
width limit analysis including information on the effects of
uncertainties in the wavelength of the band center.  If a redshifted
DIB wavelength is near a problematic region then an uncertainty in
rest wavelength can introduce a large uncertainty in a measured
equivalent width limit.  This is a more substantial issue for the
$\lambda$4428 DIB, which has the largest uncertainty in its rest-frame
wavelength ($\sim1.4$~\AA).  For the majority of the equivalent width
limits, the largest uncertainty is due to the continuum fit.

Only three DIBs have been detected in DLAs, the $\lambda$4428 DIB
reported in \citet{junk04}, and the $\lambda$5705 and $\lambda$5780
DIBs reported in a companion paper of this work by \citet{lawt06}.
All three arise in the $z=0.524$ DLA toward AO~0235+164.  The
equivalent widths were calculated by summing the individual equivalent
widths of each pixel.  Both of the detections from \citet{lawt06} were
at least one resolution element away from sky lines.  The undetected
$\lambda$5797 and $\lambda$6284 DIBs toward the $z=0.524$ AO~0235+164
DLA were located within strong sky lines.  \citet{lawt06} estimated
the equivalent width limits for these DIBs using synthetic absorption
features with varying \hbox{\small FWHMs}.  The undetected
$\lambda$6613 DIB is near a sky line; \citet{lawt06} estimated its
equivalent width limit directly from the signal-to-noise, dispersion
of the chip, and number of pixels.

\subsection{Measurements and Spectra}

The measured equivalent widths, equivalent width limits (3~$\sigma$),
and their 1~$\sigma$ uncertainties are listed in Table~\ref{tab:EWs}.
Tabulated are the DLA (by number), the corresponding QSO, the
absorption redshift of the DLA, the facility and instrument, and the
rest-frame equivalent widths or limits, with uncertainties (m{\AA}) of
each of the DIBs if they were observable.  Spectra from which the most
stringent limits were obtained are shown in
Figs.~\ref{s:0235}--\ref{s:1229_VLT}\footnote{The VLT/UVES spectrum is
not shown because only the $\lambda$4428 DIB is covered.  The UVES
spectrum has a high resolution that makes identifying the broad
$\lambda$4428 DIB very difficult.}.  Each panel ($a$)--($e$) displays
the region around a redshifted DIB.  The upper sub-panels display the
normalized flux (histogram) with the expected positions of the DIBs
(marked by ticks) based upon the {\Lya} redshift.  With the exceptions
of Figs.~\ref{s:0827} and \ref{s:1127_Gem}, the center sub-panels
display the uncertainty spectra of the associated normalized fluxes,
and the lower sub-panels display the sky counts normalized by the
continuum.  Sky data were unavailable for B2~0827+243 at $z=0.518$ and
PKS~1127--145 at $z=0.313$.

The smooth thin curves through the data are the expected observed DIB
absorption profiles based on the measured $N({\HI})$ for the DLAs,
where the band centers and \hbox{\small FWHM} of each DIB are taken
from \citet{jenn94}.  These models are not computed for the
$\lambda$4428 and $\lambda$6613 DIBs since these DIBs have no
published $N({\HI})$ relationships.  In the following sections we will
discuss limits on reddening and gas-to-dust ratios.  Thus, we also
illustrate (thick curves) the expected observed DIB absorption
profiles assuming an $E(B-V)=0.04$, the upper limit for high redshift
DLAs assuming SMC-like extinction \citep{elli05}.  For the AO~0235+164
DLA, we adopted the measured $E(B-V)=0.23$ \citep{junk04}).  The
computations of the reddening models are discussed in
\S~\ref{sec:reddening}.

The equivalent widths of the DIBs estimated from the Galactic
DIB--$N({\HI})$ and Galactic DIB--$E(B-V)$ models
\citep[see][]{welt06} are presented in Table~\ref{tab:Models}.
Columns 1--2 list the DLA number and the QSO with associated DLA
redshift.  Columns 4--8 provide the observed equivalent widths and
equivalent width limits for the $\lambda$4428, $\lambda$5780,
$\lambda$5797, $\lambda$6284, and $\lambda$6613 DIBs.  Also listed are
the predicted equivalent widths of the DIBs, where EW[$N$(HI)] denotes
the Galactic $N({\HI})$ scaling and EW[$E(B-V)$] denotes the reddening
scaling in {m\AA}.  $E(B-V)_{\rm lim}$ is the reddening upper limit for
each DLA based upon the observed DIB equivalent width limits and the
reddening relation (see \S~\ref{sec:reddening}). 

\section{Results and Discussion}

\label{sec:results}

In this section, we examine the DIB strengths in DLAs compared to the
Galactic DIB--$N({\HI})$ relation and the Galactic $E(B-V)$
relation. The former provides information on the gas content and the
latter provides information on the reddening (an indirect indicator of
dust content).  Our observations allow us to estimate lower limits on
the gas-to-dust ratios, $N({\HI})/E(B-V)$, of the DLAs in our sample.

\subsection{Gas Content}
\label{sec:gas}
The widely observed Galactic DIB--$N({\HI})$ relation describes the
correlation of the equivalent width of the $\lambda$5780,
$\lambda$5797, and $\lambda$6284 DIBs with the column density of
neutral hydrogen along the same line of sight \citep{herb95,welt06}.
\citet{welt06} extend the Galactic work by including Magellanic Cloud
sightlines.

\begin{figure}
\includegraphics[angle=0, width=1.0\textwidth]{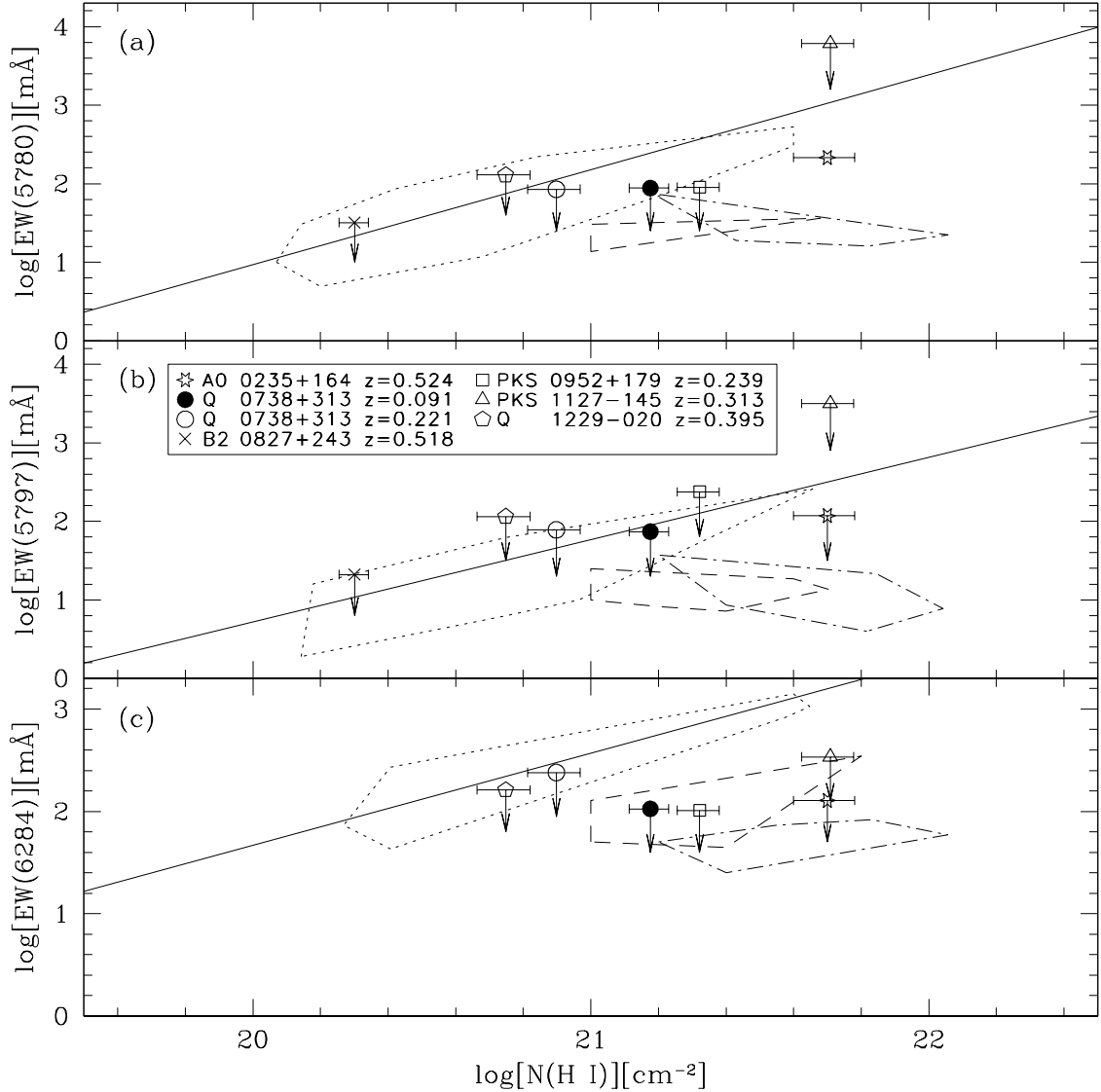}
\caption{The DIB equivalent width--$N$(HI) relations \citep{welt06}
with our DLAs added.  ---($a$) $\lambda$5780 DIB.  ---($b$)
$\lambda$5797 DIB.  ---($c$) $\lambda$6284 DIB.  The solid lines are
the best-fit weighted Galactic lines.  The region enclosed by the
dotted lines contains the Galactic data.  The region enclosed by the
dashed lines contains the LMC data.  The region enclosed by the
dot-dash lines contains the SMC data.  Error bars are 1~$\sigma$, and
upper limits are marked with arrows.  The vertical error bars for
AO~0235+164 in panel ($a$) are smaller than the point size and all
values for this DLA are from \citet{lawt06}.}\label{p:NHI_known}
\end{figure}

Plotted in Fig.~\ref{p:NHI_known} are the DIB--$N({\HI})$ relations
for the $\lambda$5780 (panel $a$), $\lambda$5797 (panel $b$), and
$\lambda$6284 DIBs (panel $c$), where $\log N{(\HI)}$ [cm$^{-2}$] is
plotted against the logarithm of the DIB equivalent widths [m\AA].
Also plotted are the (1~$\sigma$) error bars for measured values or
downward arrows representing upper limits.  The vertical error bars
for the $\lambda$5780 DIB detection in the AO~0235+164 DLA are smaller
than the point size.  The solid lines represent the weighted best-fits
to the Galactic data from \citet{welt06}.  The equivalent widths
predicted at a given $N({\HI})$, the EW[$N({\HI})$] listed in
Table~\ref{tab:Models}, are computed using this best-fit line and the
$N({\HI})$ of each DLA (Figs.~\ref{s:0235}--\ref{s:1229_VLT} contain
these models superimposed on the data).  The dotted lines in
Fig.~\ref{p:NHI_known} roughly enclose the regions containing the
Galactic data.  The dashed lines roughly enclose the regions
containing the LMC data, and the dot-dashed lines roughly enclose the
regions containing the SMC data.

The DIBs in the DLAs whose equivalent width limits in
Fig.~\ref{p:NHI_known} lie below the Galactic best-fit lines are
robust enough that we should have detected them if they followed the
same dependence on $N({\HI})$ as Galactic sightlines. In several DLAs,
the limits are inconsistent with the expectations from the Galactic
sightlines; the DIB strengths are weaker than expected.  The DIB
limits are consistent with the strengths of DIBs in the LMC or SMC.
However, higher signal-to-noise data are required to determine whether
the DIB strengths are actually consistent with or are even weaker than
those observed in the LMC and SMC.  The $\lambda$6284 DIB, panel ($c$)
of Fig.~\ref{p:NHI_known}, provides the most stringent evidence that
the DIB strengths in four DLAs are not consistent with DIB strengths
in the Milky Way.  The four DLAs toward AO~0235+164 ($z=0.524$),
Q0738+313 ($z=0.091$), PKS~0952+179 ($z=0.239$), and PKS~1127--145
($z=0.313$) are underabundant in their $\lambda$6284 DIB strengths by
factors of 4-10 times expected from the Galactic DIB--$N({\HI})$
relation.

A number of environmental factors such as reddening, metallicity, and
ionizing radiation may be responsible for DLAs not following the
Galactic DIB-$N({\HI})$ relation.  Whether it is one or a combination
of these factors, the data suggest that the environments probed by DLA
sightlines differ from those within the Milky Way in which DIBs are
present.  The environments that give rise to the DIBs are likely very
localized.  Galactic properties vary even on small scales, which is
why \citet{cox07} find varying differences and DIB strengths along
different sightlines within the larger confines of the SMC.  QSO
sightlines probe relatively small transverse spatial scales in DLA
galaxies.  Therefore, our results do not eliminate the possibility
that DIBs follow the Galactic DIB--$N({\HI})$ relationships elsewhere
in these galaxies.  We are less inclined to suggest that our results
indicate redshift evolution of organics, since infrared emission from
PAHs has been observed in the extremely dusty environments of ULIRGs
as high as redshift $z \sim 2$ \citep{yan05}.

\subsection{Reddening}
\label{sec:reddening}

There has been a long history of investigating DIB dependence on color
excess \citep{merr38, herb93}.  Whereas Galactic sight-lines along the
disk (low Galactic latitudes) have typical reddening values of 0.1 to
1.0, DLAs typically have lower reddening.  \citet{murp04} use Sloan
Digital Sky Survey (SDSS) QSO spectra to estimate $E(B-V)<0.02$ mag
(3~$\sigma$) along DLA sightlines at $z \sim 3$, assuming SMC-like
extinction.  \citet{elli05} use radio selected QSOs from the Complete
Optical and Radio Absorption Line System (CORALS) survey to estimate
reddening along DLA sightlines.  The CORALS survey avoids the
potential problem of bias from optical luminosity selected samples
that may inhibit the detection of more reddened QSOs.  They find
$E(B-V)<0.04$ (3~$\sigma$) assuming SMC-like extinction for $1.9 \leq
z \leq 3.5$ DLAs.  Using a sample of $0.8 \leq z \leq 1.3$ DLAs
selected by {\CaII} absorption, \citet{wild05} estimate an average
reddening of $E(B-V)=0.06$ assuming LMC and SMC extinction curves.

The correlations between reddening and the $\lambda$5780,
$\lambda$5797, and $\lambda$6284 DIB strengths are tight among Galactic
and Magellanic Cloud sightlines, with best fits \citep{welt06},
\begin{equation}\label{EQ:5780_reddening}
\log\mbox{EW}[E(B-V)]_{\lambda5780} = 0.99 \log E(B-V) + 2.65,
\end{equation}
\begin{equation}\label{EQ:5797_reddening}
\log\mbox{EW}[E(B-V)]_{\lambda5797} = 0.99 \log E(B-V) + 2.26,
\end{equation}
and
\begin{equation}\label{EQ:6284_reddening}
\log\mbox{EW}[E(B-V)]_{\lambda6284} = 0.80 \log E(B-V) + 3.08.
\end{equation}
The rms scatter of the measured equivalent widths about these
relationships are less than 0.15 dex; in other words, the
$\lambda$5780, $\lambda$5797, and $\lambda$6284 DIBs fairly equally
obey the DIB--$E(B-V)$ relation in the Galaxy.
Combining Galactic and extragalactic data, \citet{elli07} find a
$\lambda$5780 DIB--$E(B-V)$ relation with a slightly larger slope,
1.27, and intercept, 2.77.  Compared to \citet{welt06}, the
\citet{elli07} results yield a 10\% difference in
$\log\mbox{EW}[E(B-V)]_{\lambda5780}$ for unit reddening.

Correlations with the $\lambda4428$ DIB equivalent width are
notoriously problematic to compute because of the difficulty in
measuring the DIB's broad width.  Previous work has noted a
correlation of the $\lambda6613$ DIB equivalent width with $E(B-V)$
\citep{megi05,thor03,wese01}.  In the case of both the $\lambda$4428
and $\lambda$6613 DIBs, we employ the reddening relations provided by
T.P. Snow (unpublished, private communication).  Using the best-fit
for 75 Galactic sightlines for $\lambda4428$ and 123 Galactic
sightlines for $\lambda6613$,
\begin{equation}\label{EQ:4428_reddening}
\mbox{EW}[E(B-V)]_{\lambda4428} = 2093.99 E(B-V),
\end{equation}
and
\begin{equation}\label{EQ:6613_reddening}
\mbox{EW}[E(B-V)]_{\lambda6613} = 217.06 E(B-V).
\end{equation}
The scatter about these relations is relatively large when compared to
the other DIBs studied in this work.  The $\lambda4428$ DIB--$E(B-V)$
relation has a Pearson correlation coefficient of 0.825, while the
$\lambda6613$ DIB--$E(B-V)$ relation has a Pearson correlation
coefficient of 0.831 (unpublished, T.P. Snow, private communication).

\begin{figure}
\includegraphics[angle=-90,width=1.0\textwidth]{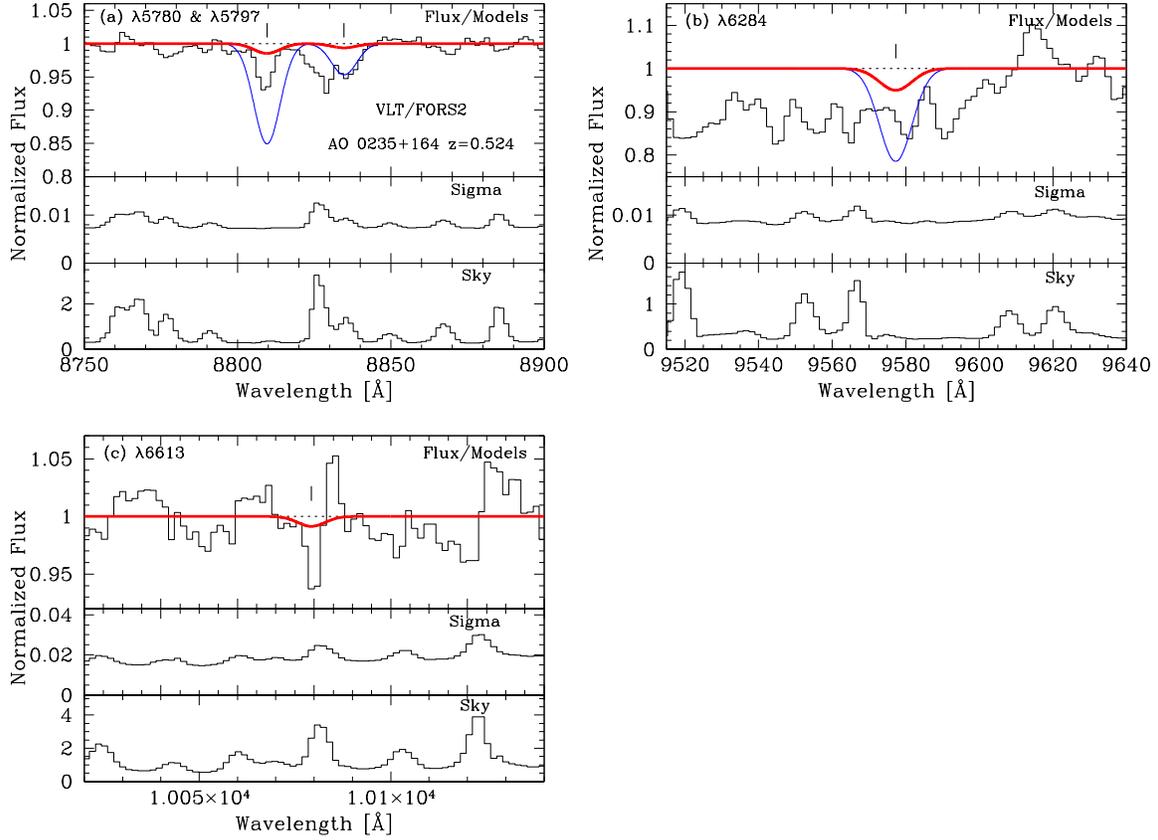}
\caption{VLT/FORS2 spectrum of the $z=0.524$ DLA toward AO~0235+164.
Plotted are the expected positions of the ($a$) $\lambda 5780/5797$,
($b$) $\lambda 6284$, and ($c$) $\lambda 6613$ DIBs.  The upper
sub-panels are the normalized flux of the data and models.  The center
sub-panels are the sigmas of the associated data fluxes normalized by
the continuum.  The lower sub-panels are the sky fluxes normalized by
the continuum.  The smooth curves are model predictions (see text).
The thin curves are the expected DIB profiles given the measured DLA
$N({\HI})$ and the Galactic DIB--$N({\HI})$ relation.  The thick
curves are the expected DIB profiles using the measured $E(B-V)=0.23$
from \citet{junk04}.\label{s:0235}}
\end{figure}

\begin{figure}
\includegraphics[angle=-90,width=1.0\textwidth]{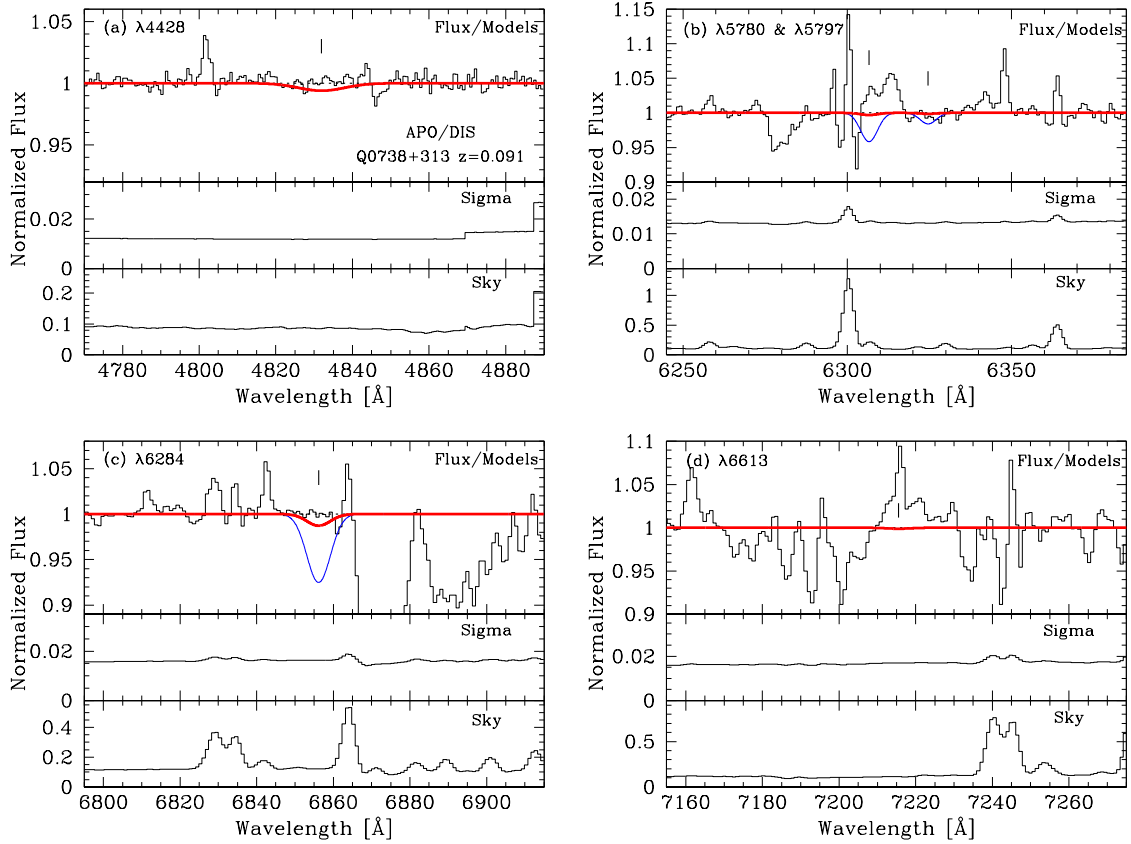}
\caption{Same as Fig.~\ref{s:0235}, but for the APO/DIS spectrum of
the $z=0.091$ DLA toward Q0738+313.  Plotted are the expected
positions of the ($a$) $\lambda 4428$, ($b$) $\lambda 5780/5797$,
($c$) $\lambda 6284$, and ($d$) $\lambda 6613$ DIBs.  The smooth
curves are model predictions (see text).  The thin curves are the
expected DIB profiles given the measured DLA $N({\HI})$ and the
Galactic DIB--$N({\HI})$ relation.  The thick curves for this DLA are
the expected DIB profiles using the $E(B-V)=0.04$ upper limit for
high-z DLAs from \citet{elli05}.\label{s:0738_091}}
\end{figure}

\begin{figure}
\includegraphics[angle=-90,width=1.0\textwidth]{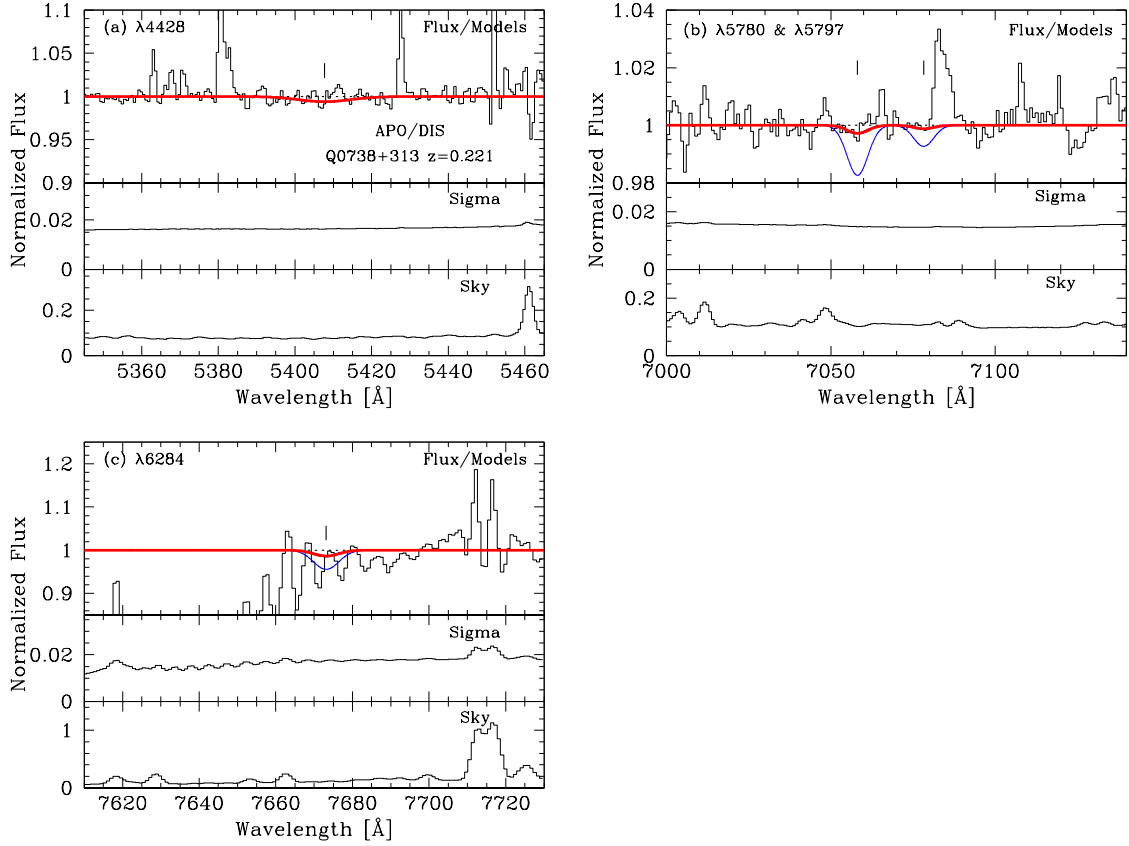}
\caption{Same as Fig.~\ref{s:0738_091}, but for the APO/DIS spectrum of
the $z=0.221$ DLA toward Q0738+313.  Plotted are the expected
positions of the ($a$) $\lambda 4428$, ($b$) $\lambda 5780/5797$, and
($c$) $\lambda 6284$ DIBs.\label{s:0738_221}}
\end{figure}

\begin{figure}
\includegraphics[angle=-90,width=1.0\textwidth]{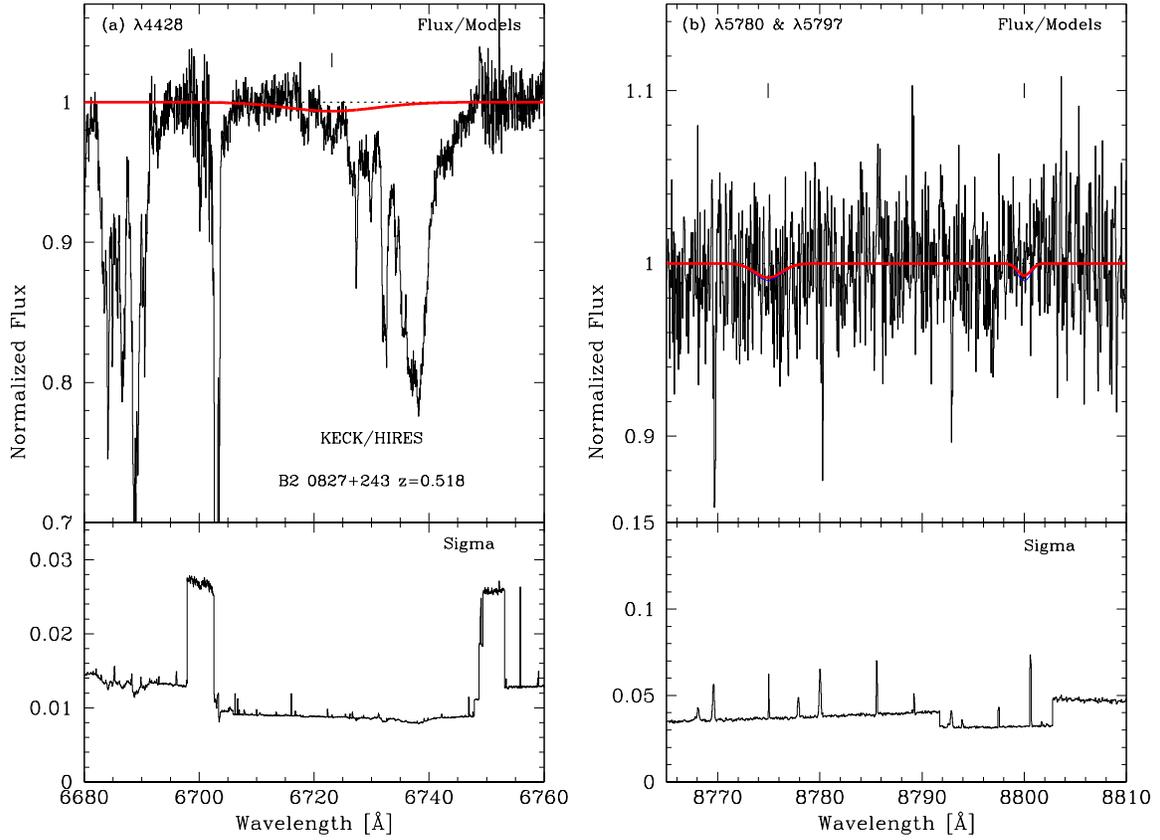}
\caption{Same as Fig.~\ref{s:0738_091}, but for the Keck/HIRES spectrum of
the $z=0.518$ DLA toward B2~0827+243.  Plotted are the expected
positions of the ($a$) $\lambda 4428$ and the ($b$) $\lambda
5780/5797$ DIBs.\label{s:0827}}
\end{figure}

\begin{figure}
\includegraphics[angle=-90,width=1.0\textwidth]{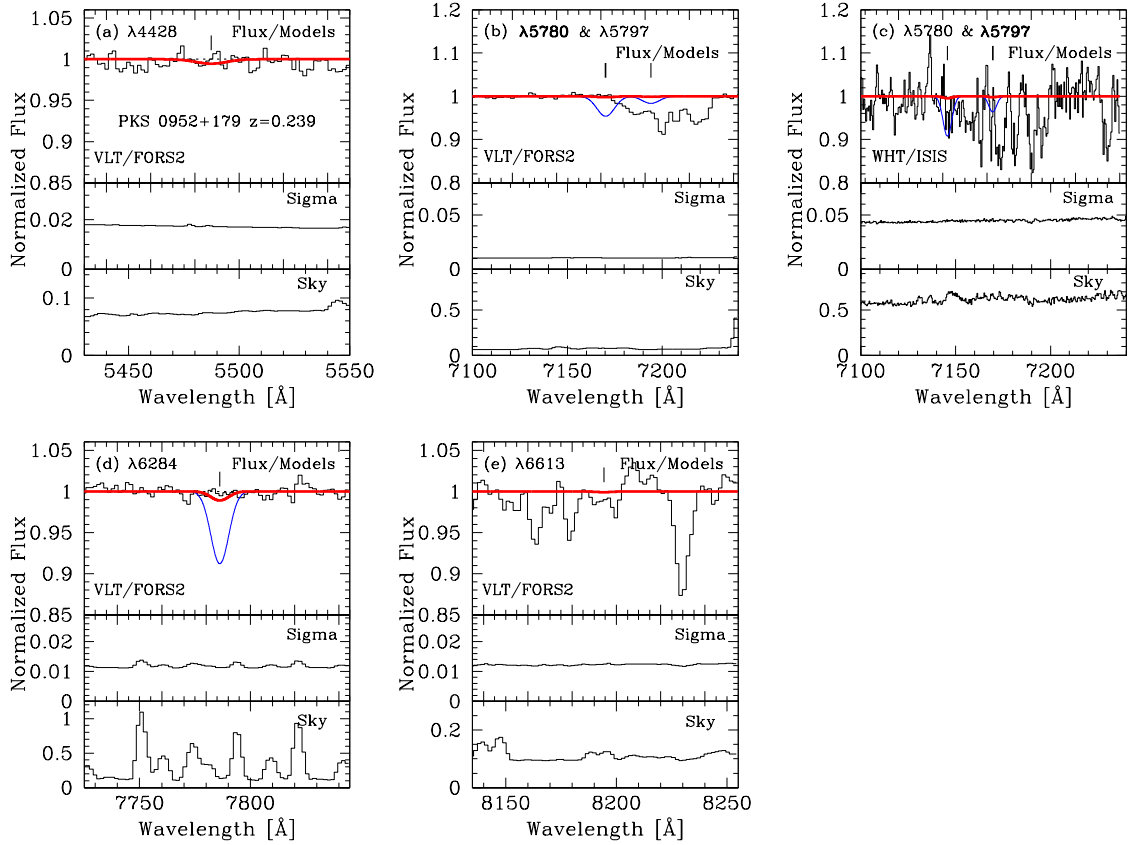}
\caption{Same as Fig.~\ref{s:0738_091}, but for the VLT/FORS2 and WHT/ISIS
spectra of the $z=0.239$ DLA toward PKS~0952+179.  Plotted are the
expected positions of the ($a$) $\lambda 4428$, ($b$) $\lambda
5780/5797$, ($c$) $\lambda 5780/5797$ (WHT/ISIS), ($d$) $\lambda
6284$, and ($e$) $\lambda 6613$ DIBs.  The WHT/ISIS spectrum provides
the adopted limit for the $\lambda$5797 DIB; the VLT/FORS2 spectrum
provides the adopted limits for the remaining DIBs.\label{s:0952_VLT}}
\end{figure}

\begin{figure}
\includegraphics[angle=-90,width=1.0\textwidth]{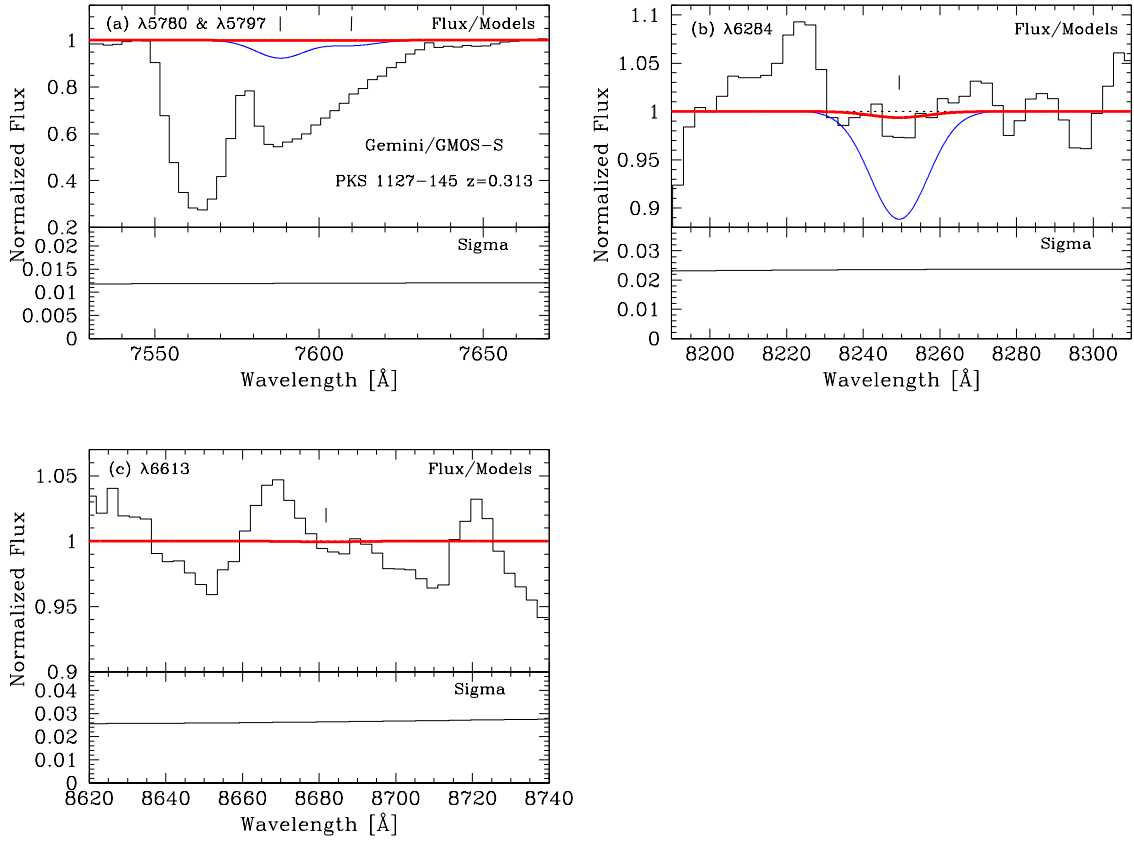}
\caption{Same as Fig.~\ref{s:0738_091}, but for the Gemini/GMOS-S spectrum
of the $z=0.313$ DLA toward PKS~1127--145.  Plotted are the expected
positions of the ($a$) $\lambda 5780/5797$, ($b$) $\lambda 6284$, and
($c$) $\lambda 6613$ DIBs.\label{s:1127_Gem}}
\end{figure}

\begin{figure}
\includegraphics[angle=-90,width=1.0\textwidth]{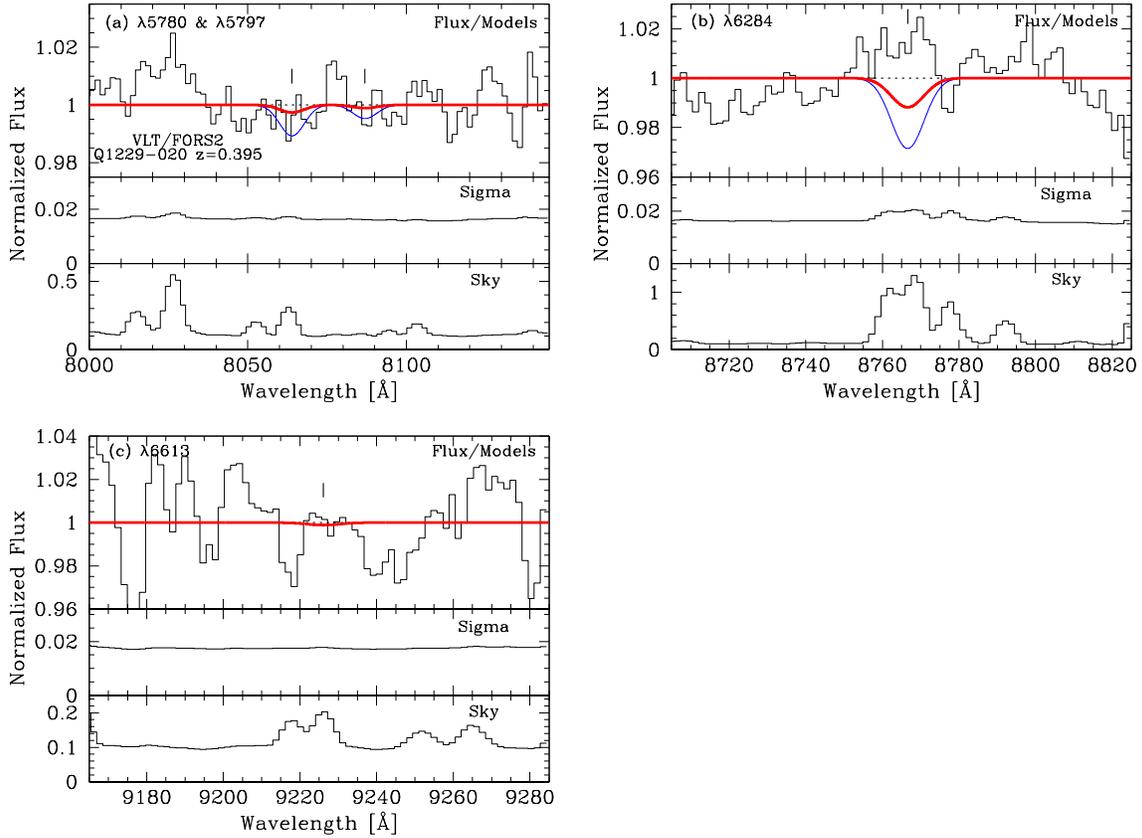}
\caption{Same as Fig.~\ref{s:0738_091}, but for the VLT/FORS2 spectrum of
the $z=0.395$ DLA toward Q1229--020.  Plotted are the expected
positions of the ($a$) $\lambda 5780/5797$, ($b$) $\lambda 6284$, and
($c$) $\lambda 6613$ DIBs.\label{s:1229_VLT}}
\end{figure}

Assuming the upper limit measured for high redshift DLAs of
$E(B-V)=0.04$ from \citet{elli05}, we computed the expected DIB
equivalent widths using Eqs.~\ref{EQ:5780_reddening},
\ref{EQ:5797_reddening}, \ref{EQ:6284_reddening},
\ref{EQ:4428_reddening}, and \ref{EQ:6613_reddening}.  The only
exception is the AO~0235+164 DLA, which has a measured reddening of
$E(B-V)=0.23$ \citep{junk04}.  The expected DIB equivalent widths are
listed in Table~\ref{tab:Models}.  Assuming these reddening values,
the expected DIB equivalent widths are systematically smaller than
those predicted by the DIB--$N({\HI})$ relationship.  In
Figs.~\ref{s:0235}--\ref{s:1229_VLT} we show the expected DIB profiles
(thick curves).  With the exception of the $\lambda6284$ DIB limit
measured from the spectrum of the DLA toward AO~0235+164 and the
$\lambda4428$ DIB limit measured from the spectrum of the DLA toward
PKS~1127--145, the expected DIB equivalent widths for adopted
reddening values are smaller than our measured equivalent width
limits; thus, if these reddening values are appropriate for
intermediate redshift DLAs, we would not have detected these DIBs.  As
discussed in \S~\ref{sec:analysis}, the $\lambda6284$ DIB limit for
the AO~0235+164 system is problematic due to a large atmospheric
feature and the $\lambda4428$ DIB limit for the PKS~1127--145 system
is unreliable due to difficulties in continuum fitting the broad DIB
within the high resolution VLT/UVES data.

Assuming the Galactic DIB--$E(B-V)$ relations are valid for DIBs in
DLAs, we used our measured equivalent width limits for the
$\lambda$4428, $\lambda$5780, $\lambda$5797, $\lambda$6284, and
$\lambda$6613 DIBs to compute upper limits on the reddening for the
DLAs in our sample directly from the weighted slopes of the
DIB--$E(B-V)$ relationships (Eqs.~\ref{EQ:5780_reddening},
\ref{EQ:5797_reddening}, \ref{EQ:6284_reddening},
\ref{EQ:4428_reddening}, and \ref{EQ:6613_reddening}) from
\citet{welt06} and T.P. Snow (unpublished, private communication). The
reddening limits, $E(B-V)_{\rm lim}$, are listed in
Table~\ref{tab:Models} for each DIB.

\begin{figure}
\includegraphics[angle=0, width=1.0\textwidth]{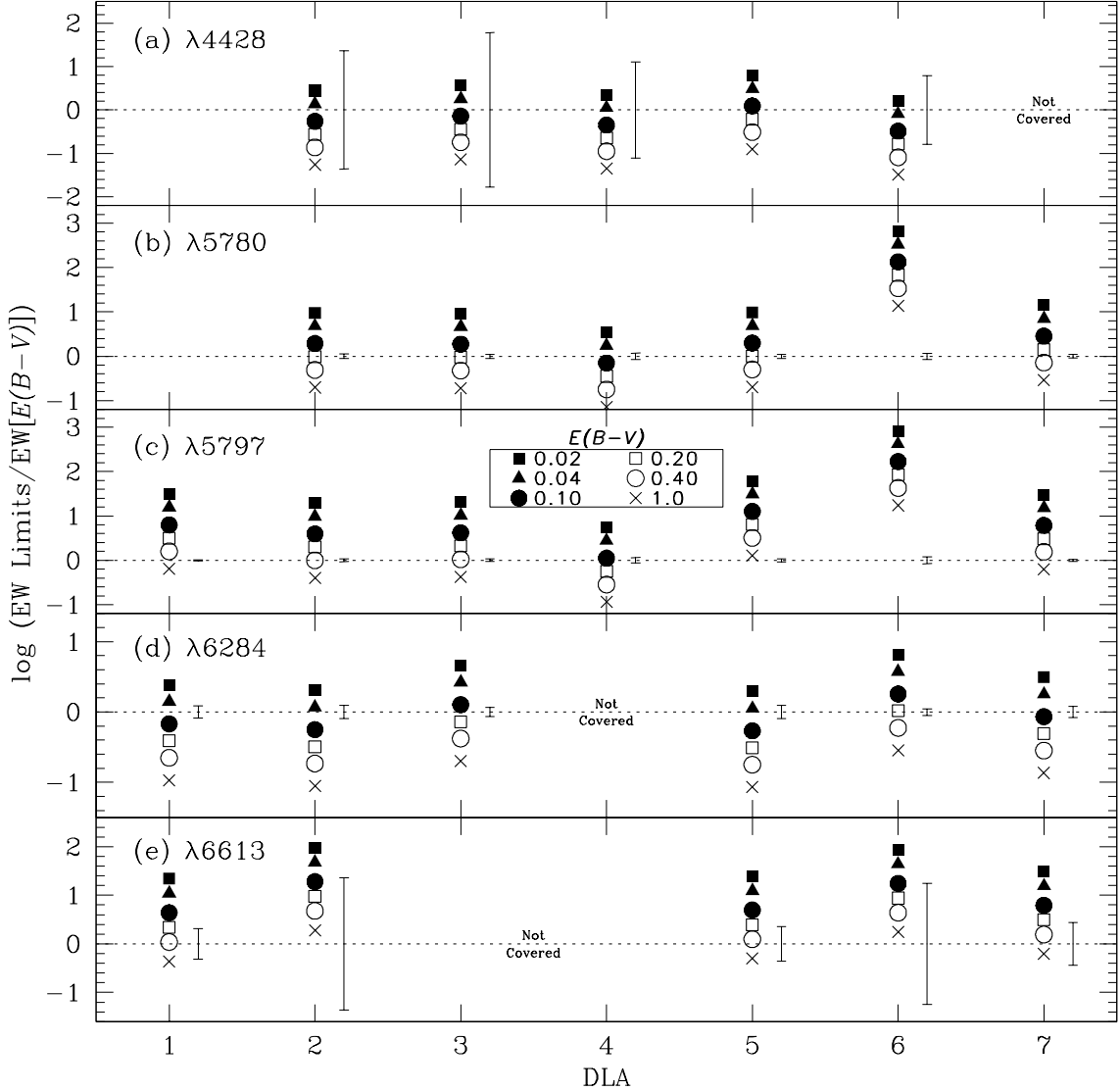}
\caption{The logarithm ratio of the measured equivalent width limits
to the equivalent width predicted for multiple values of reddening for
each DLA.  ---($a$) $\lambda$4428 DIB.  ---($b$) $\lambda$5780 DIB.
---($c$) $\lambda$5797 DIB.  ---($d$) $\lambda$6284 DIB. ---($e$)
$\lambda$6613 DIB.  The equivalent widths of the $\lambda$5780,
$\lambda$5797, and $\lambda$6284 DIBs, for each reddening, were
computed using the best-fit Galactic lines from \citet{welt06}.  The
equivalent widths of the $\lambda$4428 and $\lambda$6613 DIBs, for
each reddening, were computed from the relations in
Eqs.~\ref{EQ:4428_reddening} and \ref{EQ:6613_reddening} (unpublished,
T.P. Snow, private communication).  The DLAs labeled are as follows:
(1) AO~0235+164, (2) Q0738+313, $z=0.091$, (3) Q0738+313, $z=0.221$,
(4) B2~0827+243, (5) PKS~0952+179, (6) PKS~1127--145, (7) Q1229--020.
We leave the panels of the $\lambda$4428 and $\lambda$5780 DIBs for
AO~0235+164 blank because they are detections \citep{junk04,lawt06}
with known reddening, $E(B-V)=0.23$ \citep{junk04}.  The zero point
(horizontal dotted line) on each plot marks the estimated maximum
reddening of the DLA determined from the measured equivalent width
limit, or $E(B-V)_{\rm lim}$ (see Table~\ref{tab:Models}).  The
vertical error bars are the 1~$\sigma$ errors for the $E(B-V)_{\rm
lim}$ based on the equivalent width limit uncertainties and the
uncertainties in the DIB-reddening law slopes in
Eqs.~\ref{EQ:5780_reddening}, \ref{EQ:5797_reddening},
\ref{EQ:6284_reddening}, \ref{EQ:4428_reddening} and
\ref{EQ:6613_reddening}.  The equivalent width limits for the
$\lambda$4428 DIB measured in DLA (5) are unconstraining, thus, we
leave off the error bars in this case.  The 1~$\sigma$ errors in
panels ($b$) and ($c$) are smaller than the point
sizes.\label{p:reddening_law}}
\end{figure}

In Fig.~\ref{p:reddening_law}, we illustrate the sensitivity of this
method for $E(B-V)$ values of 0.02, 0.04, 0.10, 0.20, 0.40, and 1.0,
respectively.  Plotted are $\log (\hbox{EW Limit}/\hbox{EW}[E(B-V)])$
for each DLA for the $\lambda$4428, $\lambda$5780, $\lambda$5797,
$\lambda$6284, and $\lambda$6613 DIBs (panels $a$, $b$, $c$, $d$, $e$,
respectively).  For the condition $\log (\hbox{EW
Limit}/\hbox{EW}[E(B-V)]) \leq 0$, the reddening predicted equivalent
width for a given DIB is greater than or equal to the upper limits
afforded by our data.  A value of $\log (\hbox{EW
Limit}/\hbox{EW}[E(B-V)])=0$ corresponds to our computed reddening
upper limit, $E(B-V)_{\rm lim}$, at which the DIB should just become
detectable in our data.  The vertical error bars define the 1~$\sigma$
uncertainties in $\log (\hbox{EW Limit}/\hbox{EW}[E(B-V)])$ evaluated
at our computed $E(B-V)_{\rm lim}$.  The uncertainties are calculated
using standard error propagation taking into account the uncertainties
in both the W~Limit and the slope of the EW[$E(B-V)$] relations from
\citet{welt06} and T.P. Snow (unpublished, private communication).
The $\lambda4428$ and $\lambda5780$ DIBs for the DLA toward
AO~0235+164 are left blank because they are confirmed detections
\citep{junk04,lawt06} with a known reddening of $E(B-V)=0.23$
\citep{junk04}.

Note that in Fig.~\ref{p:reddening_law}, the $\lambda6613$
DIB--$E(B-V)$ relation is the least constraining of the DIBs in
determining the upper reddening limit, $E(B-V)_{\rm lim}$.
Furthermore, due to the scatter in the $\lambda4428$ and $\lambda6613$
DIB--$E(B-V)$ relations, the errors in the best-fit slope in
Eqs.~\ref{EQ:4428_reddening} and \ref{EQ:6613_reddening} are
relatively large which results in the large errors displayed in
Fig.~\ref{p:reddening_law}$a$ and \ref{p:reddening_law}$e$.  For these
reasons, we do not adopt upper reddening limits, $E(B-V)_{\rm lim}$,
from the $\lambda4428$ or $\lambda6613$ DIBs.

Our final adopted reddening limit for a given DLA is determined from
either the $\lambda$5780 or $\lambda$6284 DIBs.  The adopted limits
are listed in Table~\ref{tab:adoptedreddening}. Two of the DLAs have
limits of $E(B-V) \leq 0.05$ (Q0738+313 $z=0.091$ and PKS~0952+179),
both from the $\lambda$6284 DIB.  The DLA toward B2~0827+243 has an
upper limit to the reddening of $E(B-V) \leq 0.07$ from the
$\lambda$5780 DIB (the $\lambda$6284 DIB is not covered).  The
$\lambda$6284 DIB provides upper reddening limits for the remaining
DLAs toward Q1229--020 ($E(B-V) \leq 0.08$), Q0738+313 $z=0.221$
($E(B-V) \leq 0.14$), and PKS~1127--145 ($E(B-V) \leq 0.21$).  If we
were to apply constraints from the $\lambda$4428 DIB, we would infer
upper reddening limits of 0.07, 0.04, and 0.03 for the $z=0.221$ DLA
toward Q0738+313, and the DLAs toward B2~0827+243 and PKS~1127--145,
respectively.

For the DLA at $z=0.524$ toward AO~0235+164, there are two detections
from \citet{lawt06} (the $\lambda$5705 and $\lambda$5780 DIBs) and one
detection from \citet{junk04} (the $\lambda$4428 DIB).  The equivalent
widths measured from the $\lambda$4428 and $\lambda$5780 DIBs are
larger than expected from the measured $E(B-V) = 0.23$ \citep{junk04}.
The $\lambda$5797 and $\lambda$6613 DIBs give limits consistent with
this, but the $\lambda$6284 DIB suggests an $E(B-V) \leq 0.06$
\textit{by this technique}.  Essentially, we should have discovered
the $\lambda$6284 DIB assuming our technique is correct because
\citet{junk04} measures an $E(B-V)=0.23\pm0.01$.

A plausible reason for our non-detection of the $\lambda$6284 DIB in
the AO~0235+164 DLA is that the expected position of the $\lambda$6284
DIB at $z=0.524$ resides directly in a broad atmospheric absorption
band (see panel $b$ in Fig.~\ref{s:0235}).  \citet{lawt06} attempted
to overcome this by modeling the $\lambda$6284 DIB with various
equivalent widths and convolving it with the atmospheric band.  The
limit they achieve (see Table~\ref{tab:EWs}) is based on the minimum
equivalent width needed to separate the DIB out of the atmospheric
band.  This is a robust way to measure the limit.  However, there is
some inherent difficulty in this method because it has to be assumed
\textit{a priori} that the atmospheric band is not contaminated by DIB
absorption.  A second potential reason that we do not detect the
$\lambda$6284 DIB in the AO~0235+164 DLA, given its high reddening, is
that this DIB may not follow the Galactic DIB--$E(B-V)$ relation in
DLAs.  Indeed, in the Magellanic Clouds, $\lambda$6284 DIBs are weaker
than their Galactic counterparts by a factor of two relative to the
Galactic $\lambda$6284 DIB--$E(B-V)$ relation \citep{welt06}.  Given
that the $\lambda$6284 DIBs in four of our DLAs are constrained to be
4--10 times weaker than the Galactic $\lambda$6284 DIB--$N(\HI)$
relation, it is reasonable that DIB relations in DLAs could be more
Magellanic Cloud--like than Galactic--like.


Our observations have provided reddening constraints of $0.05 \leq
E(B-V) \leq 0.08$ for four DLAs at $0.09 \leq z \leq 0.52$.  This
extends to lower redshift the work of \citet{elli05}, who report
$E(B-V) \leq 0.04$ for $1.9 \leq z \leq 3.5$ DLAs, and of
\citet{murp04}, who report $E(B-V) \leq 0.02$ for $z \sim 3$ DLAs (all
limits are $3~\sigma$).  Reddening is an indirect measure of dust
content.  The slope of any DIB--$E(B-V)$ relation likely has a strong
dependence on the nature of the dust (particle size, abundance, and
composition).  \citet{junk04} argue that the dust in the AO~0235+164
DLA is more Galactic-like than Magellanic-like, but with fewer small
particles.  Because the organics responsible for the DIBs may depend
on dust for formation and/or survival, the abundance and nature of the
dust in the DLA toward AO~0235+164 might be responsible for the
presence of DIBs observed in this DLA.

\subsection{Gas-to-Dust Ratios}

Assuming that our method of estimating $E(B-V)_{\rm lim}$ realistically
reflects the upper limit on reddening in our DLAs, we can estimate
lower limits in their gas-to-dust ratios, $N({\HI})/E(B-V)$.  We
present the gas-to-dust ratio lower limits in
Table~\ref{tab:adoptedreddening}; they are computed using the measured
$N({\HI})$ from the literature and our adopted upper $E(B-V)$ limits.
The lower limits range from 2.9~$\times$~10$^{21}$ to
42~$\times$~10$^{21}$~cm$^{-2}$~mag$^{-1}$.  Note that the gas-to-dust
ratio 19.2~$\times$~10$^{21}$~cm$^{-2}$~mag$^{-1}$ for the $z=0.524$
DLA toward AO~0235+164 was measured by \citet{junk04}.

\citet{bouc85} measure lower and upper values, 37~$\times$~10$^{21}$
and 52~$\times$~10$^{21}$~cm$^{-2}$~mag$^{-1}$, respectively, for SMC
gas-to-dust ratios.  \citet{gord03} determined a gas-to-dust ratio of
19.2~$\times$~10$^{21}$~cm$^{-2}$~mag$^{-1}$ for the LMC (from their
LMC-2 data).  The average LMC sample of \citet{gord03} yields
11.1~$\times$~10$^{21}$~cm$^{-2}$~mag$^{-1}$.  A linear fit by
\citet{cox06b} to their LMC data gives a similar gas-to-dust ratio of
14.3~$\times$~10$^{21}$~cm$^{-2}$~mag$^{-1}$.  For the Milky Way,
\citet{bohl78} find a gas-to-dust ratio of
4.8~$\times$~10$^{21}$~cm$^{-2}$~mag$^{-1}$; whereas, \citet{cox06b}
find a ratio of 4.03~$\times$~10$^{21}$~cm$^{-2}$~mag$^{-1}$ from the
fit to their Galactic data.

Although the SMC, LMC, and Milky Way appear to have distinct ranges of
gas-to-dust ratios, in reality, there is a continuum that likely
reflects the variation of local environments in each.  \citet{gord03}
measure $E(B-V)$ and $N({\HI})$ toward four stars in the bar of the
SMC and one star in the wing of the SMC.  From these data, the
gas-to-dust ratios in the SMC bar span the range 17--$51 \times
10^{21}$~cm$^{-2}$~mag$^{-1}$.  These values overlap with both the LMC
gas-to-dust ratios of \citet{gord03} and \citet{cox06b} and the SMC
gas-to-dust ratios of \citet{bouc85}.  The SMC wing data yield
15.2~$\times$~10$^{21}$~cm$^{-2}$~mag$^{-1}$, which falls very near
the \citet{cox06b} LMC fit.

\citet{cox07} searched for DIBs toward six lines of sight in the bar
and wing of the SMC.  The sightline AzV456, which is in the wing of
the SMC, is the only sightline in which they detected DIBs.  From
their $E(B-V)$ and $N({\HI})$ data (Table~9 on--line material), the
gas-to-dust ratio in the SMC wing is $\sim 7$ times lower ($\sim 7.4
\times 10^{21}$ cm$^{-2}$ mag$^{-1}$) than the gas-to-dust ratio in
the SMC bar ($\sim 52.4 \times 10^{21}$ cm$^{-2}$ mag$^{-1}$).  This
SMC wing value lies between the \citet{bohl78} and \citet{cox06b}
Milky Way and the \citet{gord03} and \citet{cox06b} LMC gas-to-dust
ratios.  The SMC bar value lies just above the upper range of the
\citet{bouc85} SMC gas-to-dust ratio.  It is of interest to note that
\citet{cox07} detected DIBs only in the sightline with the smallest
gas-to-dust ratio in their sample and that this ratio approaches the
range observed for the Milky Way.

\citet{cox07} argue that the SMC wing is quiescent while the star
formation regions of the SMC bar are turbulent and exposed to larger
UV fluxes.  \citet{welt06} calculate an increase of $\sim28$--83 times
the average interstellar radiation field (ISRF) for sightlines in the
SMC bar near star-forming {\HII} regions; whereas, the SMC wing is
more similar to the Galaxy with a radiation level of $\sim0.6$ times
the average ISRF.  The variation in the gas-to-dust ratios within the
SMC are likely a reflection of the balance of dust formation
processes, such as accretion, with dust destruction \citep{cox07}.
Thus, the dust, and possibly the organics, may be destroyed in the SMC
bar explaining why \citet{cox07} could not detect DIBs in their sample
of SMC bar stars.

\begin{figure}
\includegraphics[angle=0, width=1.0\textwidth]{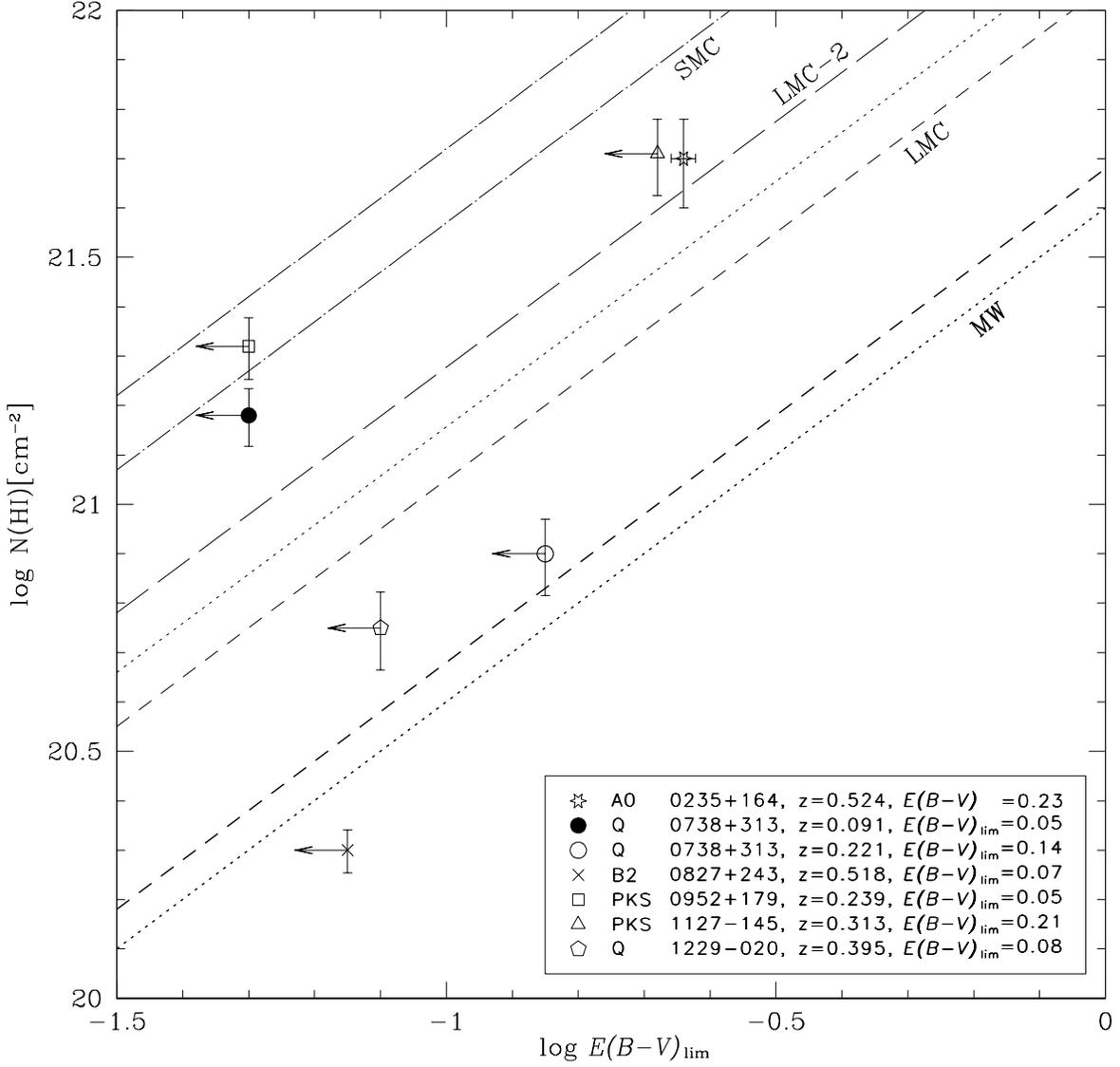}
\caption{The adopted upper limits on the gas-to-dust ratios of the
DLAs in our sample.  For comparison, the various lines provide ranges
of the measured values (given in the text) for the Milky Way (MW),
Large Magellanic Cloud (LMC), and Small Magellanic Cloud (SMC).  The
limits are illustrated with solid leftward arrows, based upon the
$\lambda$5780 and $\lambda$6284 DIBs.  The A0~0235+164 reddening
measurement (star datum point) is from
\citet{junk04}.}\label{p:Gas_Dust}
\end{figure}

In Fig.~\ref{p:Gas_Dust} we plot $\log N({\HI})$ versus $\log
E(B-V)_{\rm lim}$ for the DLAs in our sample.  The $N({\HI})$ data
points are taken from Table~\ref{tab:DLAs} and the $E(B-V)_{\rm lim}$
are taken from Table~\ref{tab:adoptedreddening}.  The $\log N({\HI})$
error bars are $1~\sigma$.  The leftward arrows represent reddening
upper limits for our sample.  For Q0738+313 at $z=0.221$, B2~0827+243,
and PKS~1127--145, the $\lambda$4428 DIB provides more stringent
limits on $E(B-V)_{\rm lim}$.  However, they should be considered less
robust due to the large errors as shown in Fig.~\ref{p:reddening_law},
panels $a$ and $e$.

Although the gas-to-dust ratios in the SMC, LMC, and Milky Way exhibit
some overlap, in general the Milky Way ISM values tend to be lower
than those of the LMC which tend to be lower than those of the SMC
\citep{bohl78,bouc85,gord03,cox06b}.  In Fig.~\ref{p:Gas_Dust}, the
dot-dashed lines illustrate the upper and lower SMC values of
\citet{bouc85}.  The middle three lines in Fig.~\ref{p:Gas_Dust}
represent gas-to-dust ratios for the LMC, and the bottom two lines
represent those for the Galaxy.  The long-dashed line gives the LMC-2
gas-to-dust ratio derived from \citet{gord03}.  The \citet{cox06b}
ratio for the LMC is given by the dot-dot line.  The short-dashed line
is the \citet{gord03} ratio from their average LMC sample.  The dashed
MW line is the \citet{bohl78} ratio, and the dotted MW line is the
\citet{cox06b} ratio.

As shown in Fig.~\ref{p:Gas_Dust}, the DLA toward PKS~0952+179 has a
gas-to-dust ratio that is consistent with or greater than the SMC.
Two of the DLAs in our sample, Q0738+313 ($z=0.091$) and
PKS~1127--145, have ratios consistent with or greater than the LMC
gas-to-dust ratios. Two, Q0738+313 ($z=0.221$) and Q1229--020, are
consistent with or greater than the Galactic ratios.  The ratio for
B2~0827+243, which has the lowest $N({\HI})$ in our sample, is not
well constrained.  The measured ratio for the $z=0.524$ DLA toward
AO~0235+164 is $N({\HI})$/$E(B-V)=19.2 \times 10^{21}$ cm$^{-2}$
mag$^{-1}$ \citep{junk04}, which is consistent with the LMC-2 ratio of
\citet{gord03} and four times larger than the Galactic ratio from
\citet{bohl78}.  This DLA has a very large $N({\HI})$, and the largest
$E(B-V)$ \citep{junk04} of the DLAs presented here.

Three DLAs are constrained to have slightly larger gas-to-dust ratios
than the $z=0.524$ DLA toward AO~0235+164, the $z=0.091$ DLA toward
Q0738+313 ($\simeq 1.5$ times greater), $z=0.239$ DLA toward
PKS~0952+179 ($\simeq 2.2$ times greater), and the $z=0.313$ DLA
toward PKS~1127--145 ($\simeq 1.2$ times greater).  However, for the
DLA toward PKS~1127--145, if we apply $E(B-V)_{\rm lim}$ from the
$\lambda$4428 DIB, the gas-to-dust ratio is $\simeq 7$ times greater
than that of the AO~0235+164 DLA and would be more consistent with
gas-to-dust ratios greater than measured in the SMC.  Again, we
caution that this more stringent ratio is not adopted due to large
errors from the scatter in the $\lambda4428$ DIB--$E(B-V)$ relation
(see Eq.~\ref{EQ:4428_reddening}).

The sightlines toward DLAs are certainly gas rich, which they share
with Galactic sightlines that contain DIBs.  However, their
environments, as probed by the background QSO, may have large
variations in $E(B-V)$, metallicity, and radiation.  Understanding the
magnitudes and properties of the dust, radiation, {\HI} and H$_{2}$
content, and metallicity in DLA samples holds promise of revealing
whether the environments of DLAs are conducive for the organics that
give rise to DIBs.  Reddening is seemingly low in the sightlines
probing the galaxies in this sample, which may play a pivotal role in
inhibiting DIB strengths.  The strengths of DIBs in DLAs do not follow
the Galactic $N({\HI})$ relation (see Fig.~\ref{p:NHI_known}).
However, the $\lambda$5780 detection in the AO~0235+164 DLA is
consistent with expectations from the DIB--$E(B-V)$ relation for
Galactic and extragalactic points \citep{elli07}.  Thus, gas-to-dust
ratio may not be as good a predictor as reddening for the presence of
DIBs in DLAs.  This would suggest that DIBs might be present in DLAs
with a wide range of gas-to-dust ratios, but that they lie within the
right-hand region of the $\log N({\HI})$--$\log E(B-V)$ plane of
Fig.~\ref{p:Gas_Dust}.

\citet{elli05} conclude that their sample of $1.9 \leq z \leq 3.5$
high redshift DLAs does not follow a Galactic gas-to-dust relation,
which is consistent with what we find for four of the DLAs in this
work (AO~0235+164 at $z=0.524$, Q0738+313 at $z=0.091$, PKS~0952+179
at $z=0.239$, and PKS~1127--145 at $z=0.313$).  However, as found by
\citet{elli05}, we cannot conclude whether the gas-to-dust ratios in
our lower redshift DLAs are consistent with the SMC gas-to-dust
ratios.  The remaining three DLAs (Q0738+313 at $z=0.221$,
B2~0827+243, and Q1229--020) do not have sufficient reddening limits
to compare with the high redshift sample of \citet{elli05}.

\subsection{Metallicity versus Reddening}

\citet{cox07} give evidence that metallicity plays an important role
in DIB strength since a high carbon abundance is necessary for the
creation of the organic molecules and the dust grains on which they
may be formed.  In Table~\ref{tab:DLAs}, we list the zinc and iron
abundances from the literature for our DLA sample.  Zinc is a good
metallicity indicator because it traces iron-group abundances and does
not readily deplete on dust.  Unfortunately, only upper limits on zinc
abundances have been measured, with the exception of the Q1229--020
DLA, where $\mbox{[Zn/H]}=-0.47$ \citep{bois98}.  Thus, we cannot
address any trends of DIB strengths with metallicity.

The Q1229--020 DLA has a metallicity roughly 0.5 dex above the average
metallicity of DLAs, and its metallicity is consistent with that of
the AO~0235+164 DLA.  Since it has a relatively stringent upper limit on
the reddening, $E(B-V)_{lim} \leq 0.08$, much below the $E(B-V) =
0.23$ of the AO~0235+164 DLA, it may be a promising candidate for
contributing to an understanding of the role of metallicity in
determining DIB strengths in DLAs.  If metallicity plays a role, it is
expected that DIB strengths in DLAs would to some degree scale with
$N(\HI)$.  The $N(\HI)$ of the Q1229--020 DLA is a full dex below that
of the AO~0235+164 DLA.  Thus, that our upper limit on the equivalent
width of the $\lambda$5780 DIB in the Q1229--020 DLA is a factor of
1.7 below that detected for the AO~0235+164 DLA is not constraining.
A very deep spectrum of the DIBs in the Q1229--020 DLA would be very
interesting.

Selecting DLAs by metal absorption would represent another approach to
searching for DIBs in external galaxies, since these DLAs may
represent a more reddened population.  \citet{wild06} propose that
{\CaII} absorbers may represent an intermediate link between the
quiescent, metal-poor and dust-poor DLAs and the intermediate
redshift, star-forming, and metal-rich Lyman break galaxies with
typical reddening of $E(B-V)=0.15$ to $0.20$.  \citet{elli07}
searched for DIBs in nine $0.07 \leq z \leq 0.55$ {\CaII}-selected
absorbers, and detected the $\lambda$5780 DIB in only one.  At this early
juncture, the {\CaII} selection success rate is comparable to that of
{\HI}-selected DLAs.  {\MgII} absorbing DLAs may also represent a
slightly more reddened population of DLAs \citep{mena07}.

\section{Conclusions}

\label{sec:conclusions}

In this paper we employ a generalized method of the \citet{schn93}
technique to find lines and determine conservative equivalent width
limits for DIBs in seven DLAs along with an assessment of
uncertainties in these limits.  We find:

(1) The $\lambda$6284 DIB in four of the DLAs in our sample have
    equivalent width upper limits that are 4--10 times lower then
    expected for similar $N({\HI})$ relative to Galactic sight lines.
    These limits are not inconsistent with the $\lambda$6284 DIB
    strengths found in the LMC and SMC.

(2) Assuming the $\lambda$5780 and $\lambda$6284 DIB--$E(B-V)$
    relations hold for DLAs, as it does for Galactic and Magellanic
    Cloud sightlines, we estimated upper limits on the reddening for
    our sample.  In four of our DLAs we estimate $E(B-V) \leq 0.08$
    with two DLAs having an $E(B-V) \leq 0.05$.  These results are
    consistent with high redshift DLA samples, which have $E(B-V) <
    0.04$ \citep{elli05} and $E(B-V) < 0.02$ \citep{murp04}.

(3) Applying the $E(B-V)$ limits, one of our DLAs is consistent with
    having the same or larger gas-to-dust ratio as the SMC. Two of our
    DLAs are consistent with having gas-to-dust ratios at least as
    large as sightlines in the LMC.  Three of our DLAs have less
    stringent limits that give lower limit gas-to-dust ratios
    consistent with Galactic, LMC, or SMC sightlines.  The AO~0235+164
    DLA has a measured $E(B-V)$ and $N({\HI})$ that puts its
    gas-to-dust fraction on the high end of the LMC sightlines as
    stated in other work \citep{junk04}.  The $E(B-V)_{lim} \leq 0.06$
    constrained from the $\lambda$6284 DIB is inconsistent with the
    known $E(B-V) = 0.23$ measured in \citet{junk04}.  We should have
    been able to detect the $\lambda$6284 DIB in the AO~0235+164 DLA
    given the limit adopted by \citet{lawt06}.  Three possibilities
    are that our method of determining the $E(B-V)_{\rm lim}$ does not
    apply to DLAs, our limits do not adequately take into account the
    large atmospheric absorption band (see Fig.~\ref{s:0235}$b$), or
    that conditions are not favorable to the formation or survival of
    this DIB carrier.

It is interesting to speculate whether ionization conditions may be an
important factor in inhibiting or enhancing DIB strengths.
\citet{welt06} test the ionization effects of DIBs along Galactic,
LMC, and SMC lines of sight.  However, they do not find any
significant trends.  On the contrary, \citet{cox07} measure the UV
radiation field along lines of sight toward the SMC and find UV
radiation is an important environmental factor in DIB strengths.
Laboratory spectroscopists claim that the DIBs may be due to partially
ionized PAHs because they produce a wealth of absorption features in
the optical spectrum \citep{snow01}.  If this is the case then a
significant UV radiation may be required.  However, UV radiation that
is too high will dissociate the molecules.  In this work, we are
unable to explore the affects of ionization conditions; knowledge of
ionization conditions in specific DLAs is difficult to obtain.

It is also interesting to speculate whether metallicity may be an
important factor in inhibiting or enhancing DIB strengths.  However,
since we have robust metallicity measurements for only two of the DLAs
in our sample, we cannot directly address the affects of metallicity.
We do point out that obtaining deep spectra of DIBs in DLAs with known
metallicity could be a fruitful future research direction, especially
if reddening were also known. A direct comparison between the low
metallicity Q1229--020 DLA and the high metallicity AO~0235+164 DLA
might be fruitful as a first examination of the affects of metallicity
in determining what governs DIB strengths in DLAs.

\subsection{Implications of our Results}

Our results imply that reddening is a more crucial indicator of DIB
strengths than is {\HI} content for DLAs; the low reddening in DLA
selected galaxies inhibits the presence of DIBs.  However, our sample
is small and additional observations are required to ascertain the
strength of this statement (metallicity and ionization conditions
likely play a role as well).

The weakness of DIB strengths in our sample hints that the
environments of high $N({\HI})$ DLA-selected galaxies may be less
suitable to create and/or sustain the organic molecules than those of
the Galaxy.  Not only is the immediate solar environment beneficial
for sustaining life, but it may be that the Galaxy is a more
hospitable location for the survival of organic molecules that may
have been the precursors to biology on Earth.  If these molecules are
important as precursors to life in the universe, charting their
presence to high redshift places constraints on how long ago and in
which environments life could have potentially formed.  The presence
of DIBs in the AO~0235+164 DLA demonstrates that organic molecules
existed in at least one DLA selected environment at a redshift of $z
\sim 0.5$, or some 5 Gyrs ago \citep{lawt06}.

Due to their weakness relative to DIBs in the Galaxy, observing DIBs
in the general population of high redshift galaxies (as opposed to
ULIRGs and star bursting galaxies) remains a challenge.  Selecting
galaxies by DLA absorption may yet prove to be a lucrative method for
detecting DIBs at high redshifts, provided the detection sensitivity
can be increased.  However, most known DLAs reside at high redshift
where the DIBs move into the near-IR, where high sensitivity
spectroscopy is time intensive.

\acknowledgments

BL acknowledges support via NASA's Graduate Student Researchers
Program (GSRP grant NNG04GN55H).  CWC acknowledges NSF grant
AST-0708210.  BAY acknowledges the support of NSERC via their
post-graduate scholarship (PGS) program, and TPS acknowledges NASA
grant NNG04GL34G for partial support of this work.  Some data were
obtained from the Gemini telescope\footnote{Based on observations
obtained at the Gemini Observatory, which is operated by the
Association of Universities for Research in Astronomy, Inc., under a
cooperative agreement with the NSF on behalf of the Gemini
partnership: the National Science Foundation (United States), the
Science and Technology Facilities Council (United Kingdom), the
National Research Council (Canada), CONICYT (Chile), the Australian
Research Council (Australia), CNPq (Brazil) and SECYT (Argentina)}
under program GS-2004A-Q-32.  We thank M. Murphy (Swinburne University
of Technology) for the VLT/UVES spectrum and W.L.W. Sargent (Caltech)
for the Keck/HIRES spectrum used in this project.  We also acknowledge
Glenn Kapcrzak (NMSU) for contributed computations.

Facilities: \facility{VLT(FORS2, UVES)}, \facility{APO(DIS)},
\facility{Keck(HIRES)}, \facility{WHT(ISIS), \facility{Gemini(GMOS)}}.

\appendix

\section{Appendix}

\subsection{Measuring Equivalent Width Detection Thresholds}

A straight forward method for determining equivalent width limits was
presented by \citet{lanz87}.  The uncertainty (or error) spectrum is
used to compute the equivalent width limit in each individual pixel,
$i$,
\begin{equation}
\sigma _{w_i} = D_i \frac{E(\lambda _i)}{I_{c}(\lambda _i)} ,
\end{equation}
where $D_i$ is the pixel dispersion in angstroms, $E(\lambda _i)$ is
the uncertainty in the flux at $\lambda _i$, and $I_{c}(\lambda _i)$
is the estimated continuum flux at $\lambda _i$.  The $1~\sigma$ equivalent width
detection threshold (limit) centered at $\lambda_i$ is then obtained
by summing the individual equivalent width limits of adjacent pixels
over a selected aperture,
\begin{equation}
\sigma _{EW} (\lambda _i) = \left[ \, \sum _{j=j_1}^{j_2} \sigma ^2_{w_j} \right] ^{1/2},
\label{EQ:ltwewlim}
\end{equation}
where $j_1 = i-m/2$, $j_2=i+m/2$, and where $m$, the aperture size, is
an even number of pixels over which the absorption feature is
anticipated.  For unresolved features, $m$ can be taken to be roughly
two resolution elements (i.e., roughly twice the number of pixels of
the full--width half maximum of an unresolved line).  Clearly,
Eq.~\ref{EQ:ltwewlim} can be generalized for resolved, broader
features by increasing $m$ to span roughly twice the full--width half
maximum of the anticipated broad line.  The aperture summation method
of \citet{lanz87} assigns equal weight to all pixels included in the
summation around $\lambda _i$.  This can result in an overestimate of
the equivalent width detection threshold of the data.

\citet{schn93} properly treat the relative weighting of pixels
adjacent to $\lambda _i$ by weighting these pixels by the instrumental
spread function, ISF.  The ISF is then a probability weighting
function of the flux centered at $\lambda _i$ with pixel weighting,
$P_j$, defined such that
\begin{equation}
\sum _{j=1}^{m} P_j = 1 ,
\end{equation}
where $m = 2j_o + 1$, and where $j_o$ is a positive integer.  For
unresolved features, the ISF, and therefore the relative values of the
$P_j$, can be taken as a Gaussian function characterized by the
Gaussian width $\sigma _{\rm ISF} = \Delta \lambda _i / 2.35$,
where $\Delta \lambda _i = \lambda _i/R$ is the full--width at half
maximum of the ISF centered at $\lambda _i$ for a spectrograph with
resolution $R$.  The value of $j_o$ is chosen appropriately for the
ISF such that $P_1 = P_m \simeq 0$ (effectively making sure that the
tails of the probability function are sampled out to where the
probabilities vanish).

The $1~\sigma$ equivalent width detection threshold for unresolved
features is then computed using
\begin{equation}
\sigma _{EW}(\lambda _i) = D_i 
\displaystyle \left[ \, \sum _{j=1}^{m} P^2_j E^2(\lambda _k)/I_{c}^{2}(\lambda _k) \right]^{1/2} 
\displaystyle \left[ \, \sum _{j=1}^{m} P^2_j \right]^{-1}  ,
\label{EQ:dpsewlim}
\end{equation}
where $k = i + (j-1) - j_o$ is the ``convolution index''.
The elements of the probability weighting function are given by
\begin{equation}
P_j = \frac{\Phi _j} {\displaystyle \sum _{j=1}^{m} \Phi _j } ,
\end{equation}
where 
\begin{equation}
\Phi _j = \exp \left\{ - \frac{ \left( \lambda_k - \lambda _i \right)
^2 }{2\sigma _{\rm ISF}^2} \right\}  .
\label{EQ:phij}
\end{equation}
If the feature is redshifted, then Eq.~\ref{EQ:dpsewlim} must be divided by
the factor $1+z$ to obtain the rest--frame equivalent width limit.

We have generalized Eq.~\ref{EQ:dpsewlim} to (1) account for resolved
features of known FWHM, and (2) account for regions in which
problematic sky--line subtraction in the vicinity of $\lambda _i$
renders the data less certain than quantified by the uncertainty
spectrum, $E(\lambda)$.  We also explicitly include redshift
dependence.

To account for resolved features, the $\sigma _{\rm ISF}$ in
Eq.~\ref{EQ:phij} is replaced with the Gaussian width of the line
spread function (LSF) of the redshifted resolved feature, which is
\begin{equation}\label{EQ:sigma}
\sigma _{\rm LSF} = \left[ \left( \frac{\hbox{\small FWHM}(1+z)}{2.35} \right) ^2 +
\sigma ^2_{\rm ISF} \right] ^{1/2} ,
\end{equation}
where FWHM is the rest--frame full--width at half maximum of the
feature, and $z$ is the redshift.  We have assumed that both the
intrinsic line shape and the ISF are well approximated by Gaussian
functions.

To account for large residuals from problematic sky--line subtraction,
the error in the flux, $E(\lambda_{k})$, in Eq.~\ref{EQ:dpsewlim} is
replaced by $\hat{E}(\lambda_k)$, which is determined by the quality
of the data at pixel $k$,
\begin{equation}\label{EQ:flux_error}
\hat{E}(\lambda_{k}) = \left\{
\begin{array} {c@{\quad \hbox{for} \quad}l}
|r(\lambda_{k})| & |r(\lambda_{k})| \geq 3 E(\lambda_{k}),\\[6pt]
E(\lambda_{k}) & |r(\lambda_{k})| < 3 E(\lambda_{k}). 
\end{array}
\right.
\end{equation}
where $|r(\lambda_{k})|$ is the residual of the flux,
$I(\lambda_{k})$, with respect to the continuum,
\begin{equation}\label{EQ:residual}
|r(\lambda_{k})| = |I(\lambda_{k}) - I_{c}(\lambda_{k})|.
\end{equation}
Using the residual gives a more conservative estimate of the
equivalent width detection threshold.  The residual of a given pixel
is used whenever the flux significantly deviates from the continuum.
In addition to poor sky--line subtraction, sources of this
deviation can be large telluric features, blending with absorption
features, or a problematic continuum fit.  Using the following
substitution for the normalized flux error simplifies the equations
for further analysis,
\begin{equation}\label{EQ:norm_flux_error}
Y(\lambda_{k}) = \frac{\hat{E}(\lambda_{k})}{I_{c}(\lambda_{k})}.
\end{equation}

Applying these criteria transforms Eq.~\ref{EQ:dpsewlim} from
\citet{schn93} into the rest--frame equivalent width detection
threshold calculation used in this work,
\begin{equation}\label{EQ:EWlimit}
\sigma_{EW}(\lambda_i) = \frac{D_{i}}{(1+z)}
\left[ \, \displaystyle \sum_{j=1}^{m} P_{j}^{2}\, Y^{2}(\lambda_{k})\right]^{1/2}
\left[ \, \displaystyle \sum_{j=1}^{m} P_{j}^{2}\right]^{-1} .
\end{equation}
The rest--frame equivalent width limits presented in
Table~\ref{tab:EWs} are quoted at the $3~\sigma$ level.

\subsection{Uncertainty Assessment}

To quantify the quality of the equivalent width limits, we estimated
the uncertainties in the $\sigma_{EW}(\lambda_i)$.  The equivalent
width limit in Eq.~\ref{EQ:EWlimit} explicitly includes the
spectrograph resolution, $R$, the absorption line FWHM, the continuum
fit, and the central wavelength of the absorption line, $\lambda_l$.
Assuming that the uncertainties in these quantities can be estimated
and are normally distributed, the variances in the equivalent width
limits are obtained from
\begin{equation}
\label{EQ:var}
V_{\sigma_{EW}} = 
  \left[\frac{\partial \sigma_{EW}}{\partial R}\delta R\right]^{2} 
+ \left[\frac{\partial \sigma_{EW}}{\partial \hbox{\small FWHM}}\delta \hbox{\small FWHM}\right]^{2} 
+ \left[\frac{\partial \sigma_{EW}}{\partial I_{c}(\lambda_{k})}\delta I_{c}(\lambda_{k})\right]^{2} 
+ \sigma_{\lambda_{l}}^{2}(\sigma_{EW}),
\end{equation}
where $\delta R$ is the uncertainty in the resolution,
$\delta\hbox{\small FWHM}$ is the uncertainty in the FWHM of the line,
$\delta I_c$ is the uncertainty in the continuum, and
$\sigma_{\lambda_{l}}(\sigma_{EW})$ is the standard deviation in the equivalent width
limit due to uncertainty in the wavelength center of the line.  The
terms in Eq.~\ref{EQ:var} for which explicit indices appear are
computed at the center pixel, $i$, of the feature.

\subsubsection{Uncertainty in Resolution}

The uncertainty in the resolution, $\delta R$, can be obtained by
measuring the full--width half maximum of unresolved sky lines and
determining the standard deviation.  Applying the chain rule, the
partial derivative of the equivalent width limit with respect to
resolution is
\begin{equation}\label{EQ:Rerror}
\frac{\partial \sigma_{EW}}{\partial R} = 
      \frac{\partial \sigma_{EW}}{\partial P_{j}}\, 
\frac{\partial P_{j}}{\partial \sigma_{\rm LSF}}\, 
\frac{\partial \sigma_{\rm LSF}}{\partial R},
\end{equation}
where,
\begin{eqnarray}
\label{EQ:parPj}
\frac{\partial \sigma_{EW}}{\partial P_{j}} &=& 
\frac{D_{i}}{(1+z)}\left[\frac{C^{1/2}}{G^{2}}\right]\left[\frac{1}{2}\frac{A\, G}{C}-B\right], \\[6pt]
\label{EQ:parsigma}
\frac{\partial P_{j}}{\partial \sigma_{\rm LSF}} &=& P_{j}\left(Q_{j} - T/S\right), \\[6pt]
\frac{\partial \sigma_{\rm LSF}}{\partial R} &=& -\frac{1}{\sigma_{\rm LSF}\, R^{3}}\left(\frac{\lambda_{i}}{2.35}\right)^{2},
\end{eqnarray}
and where,
\begin{equation}
\begin{array}{rclrcl}
\vspace{6pt}
A &=& \displaystyle\sum_{j=1}^{m}2P_{j}\, Y^{2}(\lambda_{k}) \qquad & 
Q_{j} &=& \displaystyle\frac{(\lambda_k - \lambda_i)^{2}}{\sigma_{\rm LSF}^{3}},\\\vspace{6pt}
B &=& \displaystyle\sum_{j=1}^{m}2P_{j} \qquad  & 
T     &=& \displaystyle\sum_{j=1}^{m}\frac{(\lambda_k - \lambda_i)^{2}}{\sigma_{\rm LSF}^{3}} 
          \exp \left\{-\frac{(\lambda_k - \lambda_i)^{2}}{2\sigma_{\rm LSF}^{2}}\right\},\\\vspace{6pt}
C &=& \displaystyle\sum_{j=1}^{m}P_{j}^{2}\, Y^{2}(\lambda_{k}) \qquad  &
S     &=& \displaystyle\sum_{j=1}^{m}\exp \left\{-\frac{(\lambda_k - \lambda_i)^{2}}{2\sigma_{\rm LSF}^{2}}\right\},\\
G &=& \displaystyle\sum_{j=1}^{m}P_{j}^{2}.
  & & 
\end{array}
\end{equation}
Again, the terms $P_j$ and $Q_j$ are evaluated at the center pixel,
$i$, of the feature, $j=j_o$.  Note that $Q_j$ vanishes at the line
center.

\subsubsection{Uncertainty in FWHM}

 In the case where the $\hbox{\small FWHM}$ has a known uncertainty,
 $\delta\hbox{\small FWHM}$, the affect on our equivalent width limits
 can be calculated.  Again, applying the chain rule, the partial
 derivative of the equivalent width limit with respect to FWHM is
\begin{equation}\label{EQ:FWHMerror}
\frac{\partial \sigma_{EW}}{\partial \hbox{\small FWHM}} = 
\frac{\partial \sigma_{EW}}{\partial P_{j}}\,
\frac{\partial P_{j}}{\partial \sigma_{\rm LSF}}\,
\frac{\partial \sigma_{\rm LSF}}{\partial \hbox{\small FWHM}},
\end{equation}
where $\partial \sigma_{EW}/\partial P_{j}$ and $\partial
P_{j}/\partial\sigma_{\rm LSF}$ are given by Eqs.~\ref{EQ:parPj} and
\ref{EQ:parsigma}.  From Eq.~\ref{EQ:sigma},
\begin{equation}
\frac{\partial\sigma_{\rm LSF}}{\partial \hbox{\small FWHM}} =
\left(\frac{1+z}{2.35}\right)^{2}\frac{\hbox{\small FWHM}}{\sigma_{\rm LSF}}.
\end{equation}

\subsubsection{Uncertainty in Continuum Placement}

We adopt the method of \citet{semb91} to calculate the uncertainty in
the continuum, $\delta I_{c}(\lambda_k)$, which is obtained directly
from the data by taking the rms of the residuals about the continuum,
$\sigma_c$, from $j=1$ to $m$ and multiplying by 0.5. Numerical
simulations suggest that this provides a conservative estimate of the
errors associated with the continuum \citep{semb91}.  The resulting
uncertainty is
\begin{equation}
\delta I_{c}(\lambda_{k}) = 0.5\, \sigma_{c}(\lambda_{k}),
\end{equation}
The partial derivative of the equivalent width limit with respect to
the continuum is
\begin{equation}\label{EQ:Cerror}
\frac{\partial \sigma_{EW}}{\partial I_{c}(\lambda_{k})} = \frac{1}{2}\frac{D_{i}}{(1+z)}\, G^{-1}\, U^{-1/2}\, X,
\end{equation}
where,
\begin{equation}
\begin{array}{rcl}
\vspace{6pt}
G &=& \displaystyle\sum_{j=1}^{m}P_{j}^{2},\\\vspace{6pt}
U &=& \displaystyle\sum_{j=1}^{m}P_{j}^{2}\, Y^{2}(\lambda_{k}),\\\vspace{6pt}
X &=& \displaystyle\sum_{j=1}^{m}\left\{
\begin{array} {l@{\quad \hbox{for} \quad}l}
-2P_{j}^{2}\, Y^{2}(\lambda_{k})\, I_{c}^{-1}(\lambda_{k}) & \hat{E}(\lambda_{k}) = E(\lambda_{k}),\\[3pt]
|2P_{j}^{2}\, I(\lambda_{k})\, I_{c}^{-2}(\lambda_{k})\left(1-I(\lambda_{k})\, I_{c}^{-1}(\lambda_{k})\right)| 
 & \hat{E}(\lambda_{k}) = |r(\lambda_{k})|.
\end{array}
\right.
\end{array}
\end{equation}
The $j$ elements that are selected for the computation of $X$ depend
on whether one uses the flux error, $E(\lambda_{k})$, or the residual,
$r(\lambda_{k})$, (see Eq.~\ref{EQ:residual}) for a given pixel.  The
continuum contributes a large fraction of the error in the equivalent
width limit measurements in regions where the residual is large, as
can be the case with uncorrected telluric features or problematic sky
subtraction.  In cases such as these we select portions around the
feature where the continuum estimate is reflected more accurately.

\subsubsection{Uncertainty in Central Wavelength}

If there is a known uncertainty in the wavelength center of the line,
$\delta\lambda_l$, its effect on our equivalent width limits can be
measured directly from the data.  The variance of the equivalent width
limit with respect to the wavelength center of the line is obtained by
calculating the rest equivalent width limits over the range $\Delta
\lambda_i = \pm \delta \lambda_l(1+z)$.  Eq.~\ref{EQ:EWlimit} is used
for the rest equivalent width limit calculations as before.  The
resulting variance is
\begin{equation}\label{EQ:wave_var}
\sigma_{\lambda_{l}}^{2}(\sigma_{EW}) = 
\frac{1}{N_{\sigma_{EW}}-1}
\displaystyle\sum_{n=\lambda_{i}^{-}}^{\lambda_{i}^{+}}\left(\sigma_{EW_{n}}-\left<\sigma_{EW}\right>\right)^{2},
\end{equation}
where $\lambda_{i}^{+}$ and $\lambda_{i}^{-}$ are the upper and lower
wavelengths set by $(1+z)(\lambda_l \pm \delta\lambda_{l})$,
respectively.  $N_{\sigma_{EW}}$ is the total number of equivalent
width limit calculations, and $\left<\sigma_{EW}\right>$ is the mean
equivalent width limit,
\begin{equation}\label{EQ:wave_mean}
\left<\sigma_{EW}\right> = \frac{1}{N_{\sigma_{EW}}} \displaystyle\sum_{n=\lambda_{i}^{-}}^{\lambda_{i}^{+}}\sigma_{EW_{n}}.
\end{equation}
This technique more accurately reflects the errors due to
uncertainties in wave center because it takes the actual data into
account.  If there is a problematic sky subtraction blended with the
expected position of the feature the calculation will reflect this
with a noticeably higher uncertainty.  Also, if there is an
uncertainty in the redshift, $z$, this can be incorporated into the
calculation.

\clearpage

\begin{deluxetable}{crlcclll}
\tablewidth{0pt}
\tabletypesize{\scriptsize}
\tablecaption{DLAs in Sample\label{tab:DLAs}}
\tablehead{
\colhead {} & \colhead{}  &     \colhead{}                      & \colhead{}          &
\colhead{$N$(\HI)/10$^{20}$}    &
\colhead{}       & \colhead{}  &  \colhead{}       \\
\colhead{DLA} & \colhead{}   & \colhead{QSO$\phantom{<<<}$}                   & \colhead{$z_{abs}$} &
\colhead{[atoms cm$^{-2}$]}     &         
\colhead{[Zn/H]} & \colhead{[Fe/H]} &  \colhead{Refs.}}
\tablecolumns{8}
\startdata
1 & AO  & 0235+164$^{a}$  & 0.524 & \phantom{..}50$\pm$10 &  \phantom{......}\nodata & \phantom{.......}\nodata & 1,2 \\
  & & & & & & & \\
2 & Q   & 0738+313  & 0.091 & 15$\pm$2  & $<-1.14$ & \phantom{...<}$-1.63^{+0.13}_{-0.18}$ & 3,4,8,9   \\
  & & & & & & & \\
3 &   &           & 0.221 & $\phantom{1}$7.9$\pm$1.4 &  $<-0.70^{+0.14}_{-0.17}$ & \phantom{.......}\nodata & 3,4,8,9   \\
  & & & & & & & \\
4 & B2  & 0827+243  & 0.518 & $\phantom{1}$2.0$\pm$0.2 & $<+0.30$ & \phantom{...<}$-1.02\pm0.05$ & 4,5,8,9   \\
  & & & & & & & \\
5 & PKS & 0952+179  & 0.239 & 21$\pm$3  & $<-1.02$ & \phantom{.......}\nodata & 4,5,9   \\
  & & & & & & & \\
6 & PKS & 1127--145 & 0.313 & 51$\pm$9  &  \phantom{......}\nodata & $>-2$ & 1,6,9   \\
  & & & & & & & \\
7 & Q   & 1229--020 & 0.395 & $\phantom{1}$5.6$\pm$1.0 &  \phantom{...<}$-0.47$ & $<-1.32$ & 7   \\
\enddata
\tablecomments{All upper limits are calculated to 3~$\sigma$.}
\tablenotetext{\mbox{a}}{Metallicity for AO~0235+164 is estimated via X-ray spectroscopy to be $Z=0.24\pm0.06Z_{\odot}$ \citep[\textit{Chandra},][]{turn03} and $Z=0.72\pm0.28Z_{\odot}$ \citep[\textit{ASCA \& ROSAT},][]{junk04}.}
\tablerefs{
1. Turnshek et al.\ (2003); 2. Junkkarinen et al.\ (2004); 3. Rao \& Turnshek (1998);
 4. Kulkarni et al.\ (2005); 5. Rao \& Turnshek (2000); 6. This work; 
 7. Boiss\'{e} et al.\ (1998);  8. Khare et al.\ (2004); 9. Kanekar \& Chengalur (2003)}
\end{deluxetable}

\clearpage

\begin{deluxetable}{rlccclrllrc}
\tablewidth{0pt}
\tabletypesize{\scriptsize}
\rotate
\tablecaption{Journal of Observations\label{tab:obs}}
\tablehead{
\colhead{}                 & \colhead{}            &
\colhead{}                 & \colhead{Grating/}    &
\colhead{Slit}             & \colhead{}            &
\colhead{Exposure}         & \colhead{S/N}         &  
\colhead{Wavelength}       & \colhead{}            &
\colhead{Dispersion} \\
\colhead{}                 & \colhead{QSO$\phantom{<<<}$}         &
\colhead{Facility}         & \colhead{Grism}       &
\colhead{Width}            & \colhead{Date [UT]$\phantom{<<<}$}   &
\colhead{Time [s]}         & \colhead{[pixel$^{-1}$]}             &
\colhead{Coverage [\AA]}   & \colhead{Resolution}                 &
\colhead{[\AA~pixel$^{-1}$]}}
\tablecolumns{11}
\startdata
AO  & 0235+164  & VLT/FORS2         & GRIS 600z     & 1.0''      & 2005 Jul 20       & 8400$\phantom{<}$ & 60--150 & 7318--10,744 & 1880 & 1.59 \\   
    &      &                   &               &            & 2005 Sep 6        &            &           &              & & \\
    &      &                   &               &            & 2005 Oct 1        &        &               &              & & \\
Q   & 0738+313  & APO/DIS           & HIGH          & 1.5''      & 2004 Dec 15/19    & 40,600$\phantom{<}$ & 57--84 & 4367--7817 & 2040 & 0.62/0.84 \\
    &      &                   &               &            & 2005 Feb 4        &        &                 &                  &  & (Blue/Red) \\
B2  & 0827+243  & Keck/HIRES  &               & C2/0.861'' & 1998 Dec 22       & 22,500$\phantom{<}$ & 27--114$^{a}$ & 5185--9234 & 43,000 & 0.04  \\
PKS & 0952+179  & VLT/FORS2         & GRIS 600RI    & 1.0''      & 2005 Apr 4/5/7/8  & 9000$\phantom{<}$        & 57--95 & 5298--8622   & 1650 & 1.63 \\
PKS & 0952+179  & WHT/ISIS          & R600R         & 1.0''      & 2004 May 20       & 4500$\phantom{<}$        & 20--22 & 6312--8114   & 3790 & 0.45 \\
PKS & 1127--145 & VLT/UVES    & 346,564       & 1.0''      & 2002 Jul 17/18    & 24,900$\phantom{<}$            & 38$^{b}$ & 3041--6809 & 45,000 & 0.05  \\
PKS & 1127--145 & Gemini/GMOS-S     & R400          & 1.0''      & 2004 Jun 19       & 3600$\phantom{<}$        & 38--84   & 5962--9998 & 960 & 2.75 \\ 
Q   & 1229--020 & VLT/FORS2         & GRIS 600z     & 1.0''      & 2005 Apr 9/13     & 9800$\phantom{<}$        & 50--62   & 7464--10,000 & 1860 & 1.59 \\      
Q   & 1229--020 & WHT/ISIS          & R600R         & 1.0''      & 2004 May 20       & 3600$\phantom{<}$        & 5--13   & 7212--9018    & 4980 & 0.45     
\enddata
\tablenotetext{\mbox{a}}{The low S/N is measured in the red, and the high S/N is measured in the blue.}
\tablenotetext{\mbox{b}}{The S/N is measured at the location of the $\lambda4428$ DIB.  The other DIBs are not covered in this spectrum.}
\end{deluxetable}

\clearpage

\begin{deluxetable}{crlclllllll}
\tabletypesize{\scriptsize}
\tablecolumns{11}
\tablewidth{0pt}
\rotate
\tablecaption{Equivalent Width Detections \& Limits of DIBs in DLAs\label{tab:EWs}}
\tablehead{
\colhead{}                           &
\colhead{}                           &
\colhead{}                           &
\colhead{}                           &
\colhead{}                           &
\multicolumn{6}{c}{Rest--Frame EW Detections \& Rest--Frame 3~$\sigma$ EW Limits [m\AA]}\\
\colhead{DLA}                        &
\colhead{}                           &
\colhead{QSO$\phantom{<<<}$}         & 
\colhead{$z_{abs}$}                  &
\colhead{Facility}                   &
\colhead{$\lambda$4428}         &
\colhead{$\lambda$5705}         &
\colhead{$\lambda$5780}         &
\colhead{$\lambda$5797}         &
\colhead{$\lambda$6284}         & 
\colhead{$\lambda$6613}}
\startdata
1 & AO  & 0235+164 & 0.524 & VLT/FORS2 & $\phantom{<}$741.5$\pm$26.2$^{a}$ & 63.2$\pm$8.7$^{b}$ & $\phantom{<}$216$\pm9^{b}$ & $<$118$^{bc}$ & $<$128$^{bc}$ & $<$95$^{bc}$ \\
2 & Q   & 0738+313  & 0.091 & APO/DIS  & $<$115(8) & \nodata  & $<$88(6)$^{c}$ & $<$74(3) & $<$106(3)  & $<$414(51) \\
3 &     &           & 0.221 & APO/DIS  & $<$151(11) & \nodata  & $<$85(3)  & $<$78(3) & $<$240(16)$^{c}$ & \nodata          \\
4 & B2  & 0827+243  & 0.518 & Keck/HIRES & $<$94(14)$^{a}$ & \nodata  & $<$32(1) & $<$21(1) & \nodata & \nodata \\
5 & PKS & 0952+179  & 0.239 & VLT/FORS2 & $<$259(19) & \nodata &  $<$90(4)  & $<$289(14)$^{c}$ & $<$102(4) & $<$108(6)    \\
  &     &           & 0.239 & WHT/ISIS & \nodata    & \nodata &  $<$158(4) & $<$237(18) & $<$187(3) & \nodata                   \\
6 & PKS & 1127--145 & 0.313 & VLT/UVES & $<$68(5)$^{d}$  & \nodata & \nodata & \nodata  & \nodata  & \nodata             \\
  &     &           & 0.313  & Gemini/GMOS       & \nodata  & \nodata & $<$6106(77)$^{cd}$ & $<$3162(116)$^{cd}$ & $<$341(1)$^{d}$ & $<$380(1)$^{d}$\\
7 & Q   & 1229--020 & 0.395 & VLT/FORS2 & \nodata    & \nodata & $<$129(5)  & $<$115(6) & $<$162(5) & $<$135(7)  \\
  &     &           & 0.395     & WHT/ISIS & \nodata    & \nodata & $<$292(10) & $<$198(8) & $<$627(14)$^{c}$ & \nodata  \\

\enddata
\tablenotetext{\mbox{a}}{Spectrum provided by Junkkarinen et al.\ (2004).}
\tablenotetext{\mbox{b}}{Spectrum provided by York et al.\ (2006).}
\tablenotetext{\mbox{c}}{Problematic sky subtraction or large atmospheric band.}
\tablenotetext{\mbox{d}}{Uncertainty in resolution is unknown and not included in error assessment.}
\end{deluxetable}

\clearpage

\begin{deluxetable}{crlllllll}
\tabletypesize{\scriptsize}
\tablecolumns{9}
\tablewidth{0pt}
\tablecaption{Model Predictions of DIBs in DLAs\label{tab:Models}}
\tablehead{
\colhead{}                           &
\colhead{}                           &
\colhead{QSO$\phantom{<<<}$}         &
\colhead{Observed/}                    &
\multicolumn{5}{c}{Rest--Frame W Detections, Limits (3~$\sigma$), \& Model Predictions [m\AA]}\\
\colhead{DLA}                        &
\colhead{}                           &
\colhead{$z_{\rm DLA}\phantom{<<<}$}     & 
\colhead{Models$^{c}$}               &
\colhead{$\lambda$4428}         &
\colhead{$\lambda$5780}         & 
\colhead{$\lambda$5797}         &
\colhead{$\lambda$6284}         & 
\colhead{$\lambda$6613}}
\startdata
1 & AO  & 0235+164  &  Observed & $\phantom{<}$741.5$\pm26.2^{a}$ & $\phantom{1<}$216$\pm9^{b}$   & $\phantom{.}$$<$118$^{b}$ & $\phantom{<}$$<$128$^{b}$ & $\phantom{<}$$<\phantom{.}$95$^{b}$ \\
  &     & $z=0.524$ &  EW[$N$(HI)]     &  $\phantom{<}$\nodata & $\phantom{11}$1059    & $\phantom{1..}$317     & $\phantom{11}$1572 & $\phantom{<}$\nodata\\
  &     &      &  EW[$E(B-V)=0.23$]    &  $\phantom{<}$482     & $\phantom{1<}$104     & $\phantom{<>}$43       & $\phantom{1<}$368  & $\phantom{1111}$50     \\
  &     &      &  $E(B-V)_{\rm lim}$ &  $\phantom{1111}$0.35 & $\phantom{11111}$0.48 & $\phantom{111..}$0.64  & $\phantom{11111}$0.06 & $\phantom{11111}$0.44 \\
  &     &      &                 &                       &                       &                        &            &             \\
2 & Q   & 0738+313  &  Observed         &  $<$115(8)            & $\phantom{<}$$<$88(6) & $\phantom{.1}$$<$74(3) & $\phantom{<}$$<$106(3) & $\phantom{<}$$<$414(51)  \\
  &     & $z=0.091$ &  EW[$N$(HI)]     &  $\phantom{<}$\nodata & $\phantom{1..}$247    & $\phantom{<>}$89       & $\phantom{1<}$532 & $\phantom{<}$\nodata  \\
  &     &      &  EW[$E(B-V)=0.04$]    &  $\phantom{<1}$84     & $\phantom{<>}$18      & $\phantom{1111}$8      & $\phantom{1111}$91  & $\phantom{11111}$3     \\
  &     &      &  $E(B-V)_{\rm lim}$ &  $\phantom{1111}$0.05 & $\phantom{1111}$0.20  & $\phantom{111..}$0.40  & $\phantom{11111}$0.05 & $\phantom{11111}$1.91  \\
  &     &      &                 &                       &                       &                        &            &             \\
3 & Q   & 0738+313  &  Observed         & $<$151(11)            & $\phantom{<}$$<$85(3) & $\phantom{.1}$$<$78(3) & $\phantom{<}$$<$240(16) & $\phantom{<}$\nodata     \\
  &     & $z=0.221$ &  EW[$N$(HI)]     &  $\phantom{<}$\nodata & $\phantom{1..}$114    & $\phantom{<>}$46       & $\phantom{1<}$299 & $\phantom{<}$\nodata   \\
  &     &      &  EW[$E(B-V)=0.04$]    &  $\phantom{<1}$84     & $\phantom{<>}$18      & $\phantom{1111}$8      & $\phantom{1111}$91 & $\phantom{<}$\nodata  \\
  &     &      &  $E(B-V)_{\rm lim}$ &  $\phantom{1111}$0.07 & $\phantom{1111}$0.19  & $\phantom{111..}$0.42  & $\phantom{11111}$0.14 & $\phantom{<}$\nodata\\
  &     &      &                 &                       &                       &                        &            &             \\
4 & B2  & 0827+243  &  Observed         & $\phantom{<}$$<$94(14)& $\phantom{<}$$<$32(1) & $\phantom{.1}$$<$21(1) & $\phantom{<}$\nodata    & $\phantom{<}$\nodata     \\
  &     & $z=0.518$ &  EW[$N$(HI)]     &  $\phantom{>}$\nodata & $\phantom{<>}$22      & $\phantom{<>}$11       & $\phantom{<}$\nodata    & $\phantom{<}$\nodata     \\
  &     &      &  EW[$E(B-V)=0.04$]    &  $\phantom{<>}$84     & $\phantom{<>}$18      & $\phantom{1111}$8      & $\phantom{<}$\nodata    & $\phantom{<}$\nodata     \\
  &     &      &  $E(B-V)_{\rm lim}$ &  $\phantom{1111}$0.04 & $\phantom{1111}$0.07  & $\phantom{111..}$0.11  & $\phantom{<}$\nodata    & $\phantom{<}$\nodata     \\
  &     &      &                 &                       &                       &                        &            &             \\
5 & PKS & 0952+179  &  Observed         & $<$259(19)            & $\phantom{<}$$<$90(4) & $\phantom{.}$$<$237(18)& $\phantom{<}$$<$102(4)  & $\phantom{<}$$<$108(6)   \\
  &     & $z=0.239$ &  EW[$N$(HI)]     &  $\phantom{<}$\nodata & $\phantom{1...}$370   & $\phantom{1..}$127     & $\phantom{1<}$720 & $\phantom{<}$\nodata   \\
  &     &      &  EW[$E(B-V)=0.04$]    &  $\phantom{<1}$84     & $\phantom{<>}$18      & $\phantom{1111}$8      & $\phantom{1111}$91  & $\phantom{11111}$3     \\
  &     &      &  $E(B-V)_{\rm lim}$ &  $\phantom{1111}$0.12 & $\phantom{1111}$0.20  & $\phantom{<>.}$1.29    & $\phantom{11111}$0.05 & $\phantom{11111}$0.50  \\
  &     &      &                 &                       &                       &                        &            &             \\
6 & PKS & 1127--145 &  Observed         & $\phantom{<}$$<$68(5) & $<$6106(77)           & $<$3162(116)           & $\phantom{<}$$<$341(1)  & $\phantom{<}$$<$380(1)   \\
  &     & $z=0.313$ &  EW[$N$(HI)]     & $\phantom{<}$\nodata  & $\phantom{<}$1086     & $\phantom{1..}$324     & $\phantom{11}$1601 & $\phantom{<}$\nodata  \\
  &     &      &  EW[$E(B-V)=0.04$]    &  $\phantom{<>}$84     & $\phantom{<>}$18      & $\phantom{1111}$8      & $\phantom{1111}$91 & $\phantom{11111}$3           \\
  &     &      &  $E(B-V)_{\rm lim}$ &  $\phantom{1111}$0.03 & $\phantom{<>}$14.13   & $\phantom{<>}$17.73    & $\phantom{11111}$0.21 & $\phantom{11111}$1.75     \\
  &     &      &                 &                       &                       &                        &            &             \\
7 & Q   & 1229--020 &  Observed         & $\phantom{<}$\nodata  & $\phantom{.}$$<$129(5)& $\phantom{.}$$<$115(6) & $\phantom{<}$$<$162(5)  & $\phantom{<}$$<$135(7)   \\
  &     & $z=0.395$ &  EW[$N$(HI)]     & $\phantom{<}$\nodata  & $\phantom{<>}$75      & $\phantom{<>}$32       & $\phantom{1<}$219        & $\phantom{<}$\nodata     \\
  &     &      &  EW[$E(B-V)=0.04$]    & $\phantom{<}$\nodata  & $\phantom{<>}$18      & $\phantom{1111}$8      & $\phantom{1111}$91 & $\phantom{11111}$3 \\
  &     &      &  $E(B-V)_{\rm lim}$ & $\phantom{<}$\nodata  & $\phantom{111..}$0.29 & $\phantom{111..}$0.62  & $\phantom{11111}$0.08 & $\phantom{11111}$0.62  \\
\enddata
\tablenotetext{\mbox{a}}{Spectrum provided by Junkkarinen et al.\ (2004).}
\tablenotetext{\mbox{b}}{Spectrum provided by York et al.\ (2006).}

\tablenotetext{\mbox{c}}{The limits are those that are the best
constrained from Table~\ref{tab:EWs}.  The model EW[$N$(HI)] refers to
the predicted equivalent widths (m\AA) from the Galactic best--fit
lines in Figure~\ref{p:NHI_known}.  The values used for the $N$(HI)
are taken from Table~\ref{tab:DLAs}.  The model EW[$E(B-V)=0.23$]
refers to the predicted equivalent widths (m\AA) from the Galactic
DIB--$E(B-V)$ best--fit lines with the known $E(B-V)=0.23$
\citep{junk04}.  The model EW[$E(B-V)=0.04$] refers to the predicted
equivalent widths (m\AA) from the Galactic DIB--$E(B-V)$ best--fit
lines with the upper limit of $E(B-V)=0.04$ as measured in high-z DLAs
by \citet{elli05}.  For all systems, the $E(B-V)=0.04$ except for
AO~0235+164 which has a measured $E(B-V)=0.23$ \citep{junk04}.  The
value $E(B-V)_{\rm lim}$ is the $E(B-V)$ upper limit inferred from our
equivalent width limits and the Galactic DIB--$E(B-V)$ best-fit lines
for the $\lambda$5780, $\lambda$5797, and $\lambda$6284 DIBs
\citep{welt06} and the $\lambda$4428 and $\lambda$6613 DIBs
(unpublished, T.P. Snow, private communication).}
\end{deluxetable}

\clearpage

\begin{deluxetable}{crlccrc}
\tabletypesize{\scriptsize}
\tablecolumns{8}
\tablewidth{0pt}
\tablecaption{Adopted Reddening and Gas-to-Dust Ratios\label{tab:adoptedreddening}}
\tablehead{
\colhead{DLA}                        &
\colhead{}                           &
\colhead{QSO$\phantom{<<<}$}         & 
\colhead{$z_{abs}$}                  &
\colhead{$E(B-V)$}             &
\colhead{gas/dust}                   &
\colhead{Contraint}                        \\
\colhead{}                        &
\colhead{}                           &
\colhead{}         & 
\colhead{}                  &
\colhead{[mag]}             &
\colhead{[{\cmsq}~mag$^{-1}$]}                   &
\colhead{DIB}                        
}                       
\startdata
1 & AO  & 0235+164  & 0.524\tablenotemark{\mbox{a}} &  \phantom{$<$}$0.23$  & 19.2$ \times10^{21}$ & \nodata \\
2 & Q   & 0738+313  & 0.091 &  $<0.05$  & $>$ 30\phantom{.0}$ \times10^{21}$ & $\lambda$6284 \\
3 &     &           & 0.221 &  $<0.14$  & $>$ \phantom{1}5.6$ \times10^{21}$ & $\lambda$6284 \\
4 & B2  & 0827+243  & 0.518 &  $<0.07$  & $>$ \phantom{1}2.9$ \times10^{21}$ & $\lambda$5780 \\
5 & PKS & 0952+179  & 0.239 &  $<0.05$  & $>42$\phantom{.0}$ \times10^{21}$ & $\lambda$6284 \\
6 & PKS & 1127--145 & 0.313 &  $<0.21$  & $>24$\phantom{.0}$ \times10^{21}$ & $\lambda$6284 \\
7 & Q   & 1229--020 & 0.395 &  $<0.08$  & $>$ \phantom{1}7.0$\times10^{21}$ & $\lambda$6284 \\
\enddata
\tablenotetext{\mbox{a}}{Values quoted from Junkkarinen et al.\ (2004).}
\end{deluxetable}

\clearpage

\end{document}